%% file: 0--main-acmsmall-conf.tex
\documentclass[acmsmall]{acmart}
\AtBeginDocument{%
  }

\setcopyright{cc}
\setcctype{by}
\acmDOI{10.1145/3729348}
\acmYear{2025}
\acmJournal{PACMSE}
\acmVolume{2}
\acmNumber{FSE}
\acmArticle{FSE078}
\acmMonth{7}
\received{2025-02-26}
\received[accepted]{2025-04-01}

\input{0-0-packages_commands}

\begin{document}

\title{Multi-modal Traffic Scenario Generation for Autonomous Driving System Testing}

\author{Zhi Tu}
\orcid{0000-0001-6175-164X}
\affiliation{%
  \institution{Purdue University}
  \city{West Lafayette}
  \country{USA}
}
\email{tu85@purdue.edu}

\author{Liangkun Niu}
\orcid{0009-0005-4435-8148}
\affiliation{%
  \institution{Purdue University}
  \city{West Lafayette}
  \country{USA}
}
\email{niu61@purdue.edu}

\author{Wei Fan}
\orcid{0009-0007-7249-2453}
\affiliation{%
  \institution{Purdue University}
  \city{West Lafayette}
  \country{USA}
}
\email{fan370@purdue.edu}

\author{Tianyi Zhang}
\orcid{0000-0002-5468-9347}
\affiliation{%
  \institution{Purdue University}
  \city{West Lafayette}
  \country{USA}
}
\email{tianyi@purdue.edu}

% \renewcommand{\shortauthors}{Tu et al.}

% \begin{abstract}
%   A clear and well-documented \LaTeX\ document is presented as an
%   article formatted for publication by ACM in a conference proceedings
%   or journal publication. Based on the ``acmart'' document class, this
%   article presents and explains many of the common variations, as well
%   as many of the formatting elements an author may use in the
%   preparation of the documentation of their work.
% \end{abstract}
\input{0-Abstract}


% \keywords{Do, Not, Us, This, Code, Put, the, Correct, Terms, for, Your, Paper}


% \received[accepted]{5 June 2025}

\maketitle

\input{1-Introduction}

\input{2-Approach}

\input{3-Experiments}

\input{4-Discussion}

\input{5-Related_Work}

\input{6-Conclusion}

\begin{acks}
\revisecr{We would like to thank the anonymous reviewers for their valuable feedback. 
This work is partly supported by the National Science Foundation under the grant IIS-2416835.
%% This material is based upon work supported by the National Science Foundation under Award Number 2416835.
%% Any opinions, findings, and conclusions or recommendations expressed in this material are those of the author(s) and do not necessarily reflect the views of the National Science Foundation.
}
\end{acks}

\bibliographystyle{ACM-Reference-Format}
\bibliography{0-0-bib}

% \appendix

\end{document}

%% file: 0-0-packages_commands.tex
\usepackage{graphicx}
\usepackage{textcomp}
\usepackage{xcolor}
\usepackage{subcaption}
\usepackage{colortbl}
\usepackage{multirow}
\usepackage{wrapfig}

\usepackage{enumitem}

\usepackage{listings}

\definecolor{codegreen}{rgb}{0.13,0.41,0.24}
\definecolor{codegray}{rgb}{0.5,0.5,0.5}
\definecolor{codepurple}{rgb}{0.52,0.03,0.40}
\definecolor{backcolour}{rgb}{1.0,1.0,1.0}

\lstdefinestyle{code}{ %
  columns=fixed,
  basewidth=0.53em,
  basicstyle=\ttfamily\scriptsize,
  commentstyle=\color{codegreen},
  keywordstyle=\color{codepurple},
  numberstyle=\tiny\color{codegray},
  stringstyle=\color{magenta},
  numbers=left,
  numbersep=1em,
  xleftmargin=1.5em,
  xrightmargin=0em,
  breaklines=true,
  frame=tb,
  framexleftmargin=0em,
  framexrightmargin=0em,
  rulesepcolor=\color{gray},
}

\usepackage{algorithm}
\usepackage{amsmath}
\usepackage[T1]{fontenc}

\newcommand{\todo}[1]{#1}

\newcommand{\revisepara}[1]{#1}  % doesn't work with new lines.
\newcommand{\reviseminor}[1]{#1}  
\newcommand{\revisecr}[1]{#1}  % Revision made for camera ready.
\newcommand{\tool}{\textsc{TrafficComposer}}
\newcommand{\tooltext}{\tool$_\text{text}$}
\newcommand{\toolvisual}{\tool$_\text{visual}$}
\newcommand{\textualextractor}{{textual description parser}}
\newcommand{\visualextractor}{{visual information extractor}}
\newcommand{\numads}{six}
\newcommand{\numscenario}{$120$}
\newcommand{\ratiobug}{$33\%$--$124\%$}
\newcommand{\ratiotime}{$28\%$--$58\%$}
\newcommand{\drivefuzzNumSeeds}{$40$}
\newcommand{\result}[1]{#1}
\newcommand{\commentout}[1]{}
\newcommand{\numllmfamily}{five}
\newcommand{\numllm}{eight}
\newcommand{\numobjdet}{nine}
\newcommand{\accyolo}{$98.7\%$}
\newcommand{\numdirectbug}{$37$}
\newcommand{\numrepeatllmexp}{three}
\newcommand{\rqacceff}{RQ3}
\newcommand{\rqablation}{RQ4}
\newcommand{\rqrobustllm}{RQ5}
\newcommand{\rqrobustcv}{RQ6}

%% file: 0-Abstract.tex
\begin{abstract}
Autonomous driving systems (ADS) require extensive testing and validation before deployment. 
However, it is tedious and time-consuming to construct traffic scenarios for ADS testing.
In this paper, we propose {\tool}, a multi-modal traffic scenario construction approach for ADS testing. 
{\tool} takes as input a natural language (NL) description of a desired traffic scenario and a complementary traffic scene image.  
Then, it generates the corresponding traffic scenario in a simulator, such as CARLA and LGSVL. 
Specifically, {\tool} integrates high-level dynamic information about the traffic scenario from the NL description and intricate details about the surrounding vehicles, pedestrians, and the road network from the image. 
The information from the two modalities is complementary to each other and helps generate high-quality traffic scenarios for ADS testing.
On a benchmark of {\numscenario} traffic scenarios, {\tool} achieves \result{$97.0\%$} accuracy, outperforming the best-performing baseline by \result{$7.3\%$}.
Both direct testing and fuzz testing experiments on {\numads} ADSs prove the bug detection capabilities of the traffic scenarios generated by {\tool}. 
These scenarios can directly discover {\numdirectbug} bugs and help two fuzzing methods find {\ratiobug} more bugs serving as initial seeds.
\end{abstract}

%% file: 1-Introduction.tex
\section{Introduction}
Despite significant advancements in autonomous driving, traffic accidents caused by autonomous vehicles continue to occur~\cite{apn2020Tesla,bbc2019Tesla,bbc2016Google,bbc2016Uber,tmn2018Tesla}. 
This highlights the critical need for effective testing methods for autonomous driving systems (ADSs).
A common practice in ADS testing is asking ADS developers to manually craft test scenarios, in which the developers specify the maneuvers of surrounding vehicles and the ego vehicle~\cite{carlachallenge, DBLP:journals/corr/abs-2012-10672}.
However, these manual approaches are often tedious and time-consuming.
Recently, researchers have proposed search-based testing~\cite{DBLP:conf/kbse/AbdessalemPNBS18,DBLP:conf/icse/GambiMF19, DBLP:conf/issta/GambiMF19,DBLP:conf/kbse/LuoZAJZIW021} and fuzzing~\cite{DBLP:conf/kbse/0008P00Y22, avfuzzer, DBLP:conf/ccs/KimLRJ0K22, DBLP:journals/tse/ZhongKR23, 10.1145/3597926.3598072} approaches, which employ heuristic functions or machine learning models to generate test scenarios. In particular, LawBreaker~\cite{DBLP:conf/kbse/0008P00Y22} shows that ADS fuzzing can be significantly enhanced with high-quality traffic scenarios and oracles.
However, LawBreaker still requires developers to manually specify the scenarios and oracles in their DSL.

To reduce the manual effort, we propose {\tool}, a multi-modal approach to automate the generation of executable traffic scenarios in simulation.
{\tool} uses both textual descriptions and reference images as inputs. 
The two modalities provide complementary information to describe the designated traffic scenario.
Textual descriptions provide a high-level overview of the desired scenario, \revisecr{but can become cumbersome to convey details, e.g., the specific location and color of each vehicle}.
In contrast, reference images can provide the specific details lacking in textual descriptions. 
However, due to their static nature, images fail to convey dynamic information, such as the speed and behaviors of vehicles and pedestrians within the scenario, which textual descriptions can help with.

A key feature of {\tool} is its ability to extract and align the information from multi-modal inputs. 
{\tool} leverages Large Language Models (LLMs) to efficiently extract information from textual description inputs. 
It also uses pre-trained computer vision models to analyze the reference image for the desired traffic scene.
To encapsulate the information extracted from both textual and visual inputs, {\tool} employs an intermediate representation (IR) based on a domain-specific language. 
{\tool} then aligns and merges the two IRs to generate detailed and accurate traffic scenarios, which will then be converted to an executable traffic scenario in a simulator.

To evaluate {\tool}, we first measure how closely the generated traffic scenarios \revisecr{align} with the text and image descriptions (Section~\ref{sec:setup_info_acc}). We constructed a benchmark \revisecr{of} {\numscenario} scenarios (Section~\ref{sec:exp_benchmark}) and compared {\tool} with a state-of-the-art approach, TARGET~\cite{deng2023target} \revisecr{and six multi-modal LLMs.} 
{\tool} achieved \result{$97.0\%$} accuracy, \revisecr{outperforming TARGET by an absolute margin of $22.7\%$ and the best-performing multi-modal LLM, GPT-4o, by $7.3\%$.}
%% while TARGET achieved \result{$74.3\%$} accuracy and the GPT-4o approach achieved \result{$89.7\%$} accuracy.
We then measure the bug detection capabilities of \revisecr{{\tool}-generated scenarios} by using them as \revisecr{direct} test cases and as initial seeds \revisecr{for} two ADS fuzzing methods~\cite{DBLP:conf/ccs/KimLRJ0K22, DBLP:conf/kbse/0008P00Y22} (Section~\ref{sec:setup_fuzz}). 
\revisecr{The generated scenarios} exposed {\numdirectbug} ADS \revisecr{failures} in direct testing and helped the fuzzers discover {\ratiobug} more \revisecr{ADS failures} \revisecr{compared to using the same number of original seeds}.
%% when using the same number of traffic scenarios as initial seeds as the original settings.
\revisecr{We further examine the correlation between scenario accuracy and bug-exposing effectiveness by a comparative study (Section~\ref{sec:setup_acceff}) and demonstrate the complementary contributions of textual and visual modalities via an ablation study (Section~\ref{sec:setup_modality}).
We also analyze {\tool}'s sensitivity to {\numllm} LLMs and ten computer vision models, along with the potential errors they introduce (Section~\ref{sec:setup_llm} and \ref{sec:setup_robustcv}).}
% We also demonstrated {\tool}'s information extraction capability is generalizable across {\numllm} different LLMs (Section~\ref{sec:setup_llm}).

Overall, this work makes the following contributions: 
\begin{itemize}
    \item We propose a novel multi-modal approach that automatically constructs traffic scenarios in simulation.
    \item We develop the first multi-modal benchmark for traffic scenario generation and ADS testing.
    \item We conduct a comprehensive evaluation of {\tool} with {\numscenario} scenarios and 6 autonomous driving systems. We demonstrated the bug detection capability of the traffic scenarios generated by {\tool} via both direct testing and fuzz testing. We demonstrated the contribution of each modality via an ablation study and the generalizability of {\tool} \revisecr{across {\numllm} LLMs and ten computer vision models.}
    % \item The source code and data are publicly available for replication and reproduction~\cite{my_artifacts}.
\end{itemize}

\begin{figure}[htbp]
    \centering
    \begin{subfigure}[b]{0.45\linewidth}
        \centering
        \includegraphics[width=\linewidth]{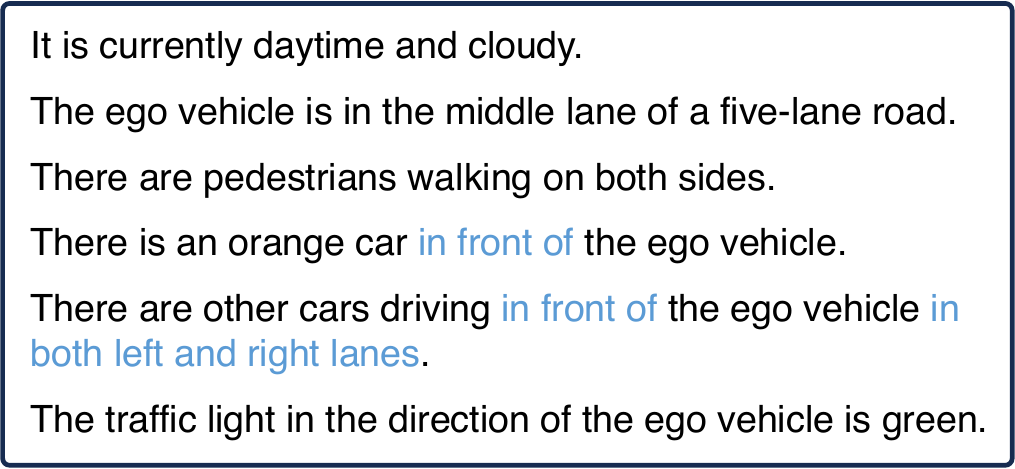}
        \caption{Textual input example.}
        \label{fig:motivation_text}
    \end{subfigure}
    % \hfill % Optional: add some horizontal separation between the subfigures
    \begin{subfigure}[b]{0.1\linewidth}
    \end{subfigure}
    \begin{subfigure}[b]{0.45\linewidth}
        \centering
        \includegraphics[width=\linewidth]{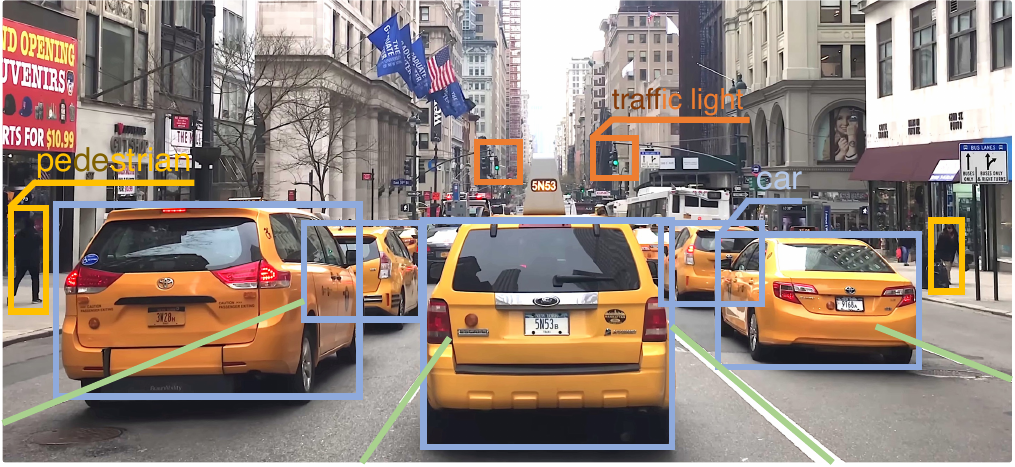}
        \caption{Visual input example.}
        \label{fig:motivation_visual}
    \end{subfigure}
    \caption{Example of two modalities of inputs of {\tool} to generate a traffic scenario simulation.}
    \label{fig:motivation}
\end{figure}
\section{Motivating Example} 
To illustrate the challenges in traffic scenario design for ADS testing, we show a motivating example in Figure~\ref{fig:motivation}.
Suppose that Alice is an ADS developer who wants to test an ADS in a particular scenario \revisecr{of interest}.
She can describe the scenario with text as in Figure~\ref{fig:motivation_text}.
In her textual description, she can efficiently convey information such as weather and time of day about the scenario.
However, she finds it cumbersome to specify the positions of multiple vehicles and pedestrians due to the inherent ambiguity of the natural language description.
For instance, Alice may write, ``\textit{there are other cars in front of the ego vehicle in both left and right lanes.}'' Yet, this description is too vague to specify the number of vehicles \revisecr{in} each lane and their exact locations, which \revisecr{are essential} for \revisecr{rendering the traffic scenario in simulators.}

Alternatively, Alice can use a reference image to enhance the description of the traffic scenario, as in Figure~\ref{fig:motivation_visual}.
This image clearly depicts the lane divisions, the precise positions of surrounding vehicles and pedestrians, and the status of the traffic lights.
With this information, the desired traffic \revisecr{scenario becomes} more detailed.
However, the reference image does not capture the dynamic aspects of the traffic \revisecr{scenario}, such as vehicle speeds and maneuvers.
Therefore, Alice should employ \revisecr{both the textual description and the reference image} as complementary inputs, combining dynamic and detailed spatial information for a comprehensive description.

Existing end-to-end traffic scenario generation methods, such as AC3R~\cite{DBLP:conf/sigsoft/GambiHF19} and RMT~\cite{DBLP:journals/corr/abs-2012-10672} face challenges in accommodating free-form traffic scenario descriptions.
AC3R is designed to reconstruct car crash scenarios from police reports. 
It requires input descriptions in a detailed and non-ambiguous narrative.
%% , as in a police report. 
For instance, all vehicles in the traffic \revisecr{scenario} must be explicitly described with their make, model, color, etc.
Similarly, RMT requires input descriptions to strictly follow the ``\textit{if--then}'' grammar.
TARGET~\cite{deng2023target} supports a more flexible input format and can process traffic law descriptions in driving handbooks\reviseminor{.}
However, it exclusively relies on textual input and thus suffers from the inherent ambiguity of natural language descriptions. Consequently, the traffic scenarios generated by TARGET are relatively simple, typically involving only one vehicle or pedestrian.
Domain-specific languages like Scenic~\cite{fremont2019scenic} can be used to generate complex traffic scenarios. However, using such DSLs demands extensive manual effort and proficiency from developers. 
For example, specifying our motivating example scenario in Scenic requires $114$ lines of manual coding, as shown in our source code repository~\cite{my_artifacts_motivating_example_scenic}.

Our 
% proposed 
approach, {\tool}, overcomes these challenges by automatically extracting and aligning information from two complementary modalities, enhancing scenario design precision and comprehensiveness.
Additionally, {\tool} automates the generation of executable traffic scenarios in simulation for ADS testing, enabling efficient traffic scenario design and generation.

%% file: 2-Approach.tex
\begin{figure}[t]
    \centering
    \includegraphics[width=\linewidth]{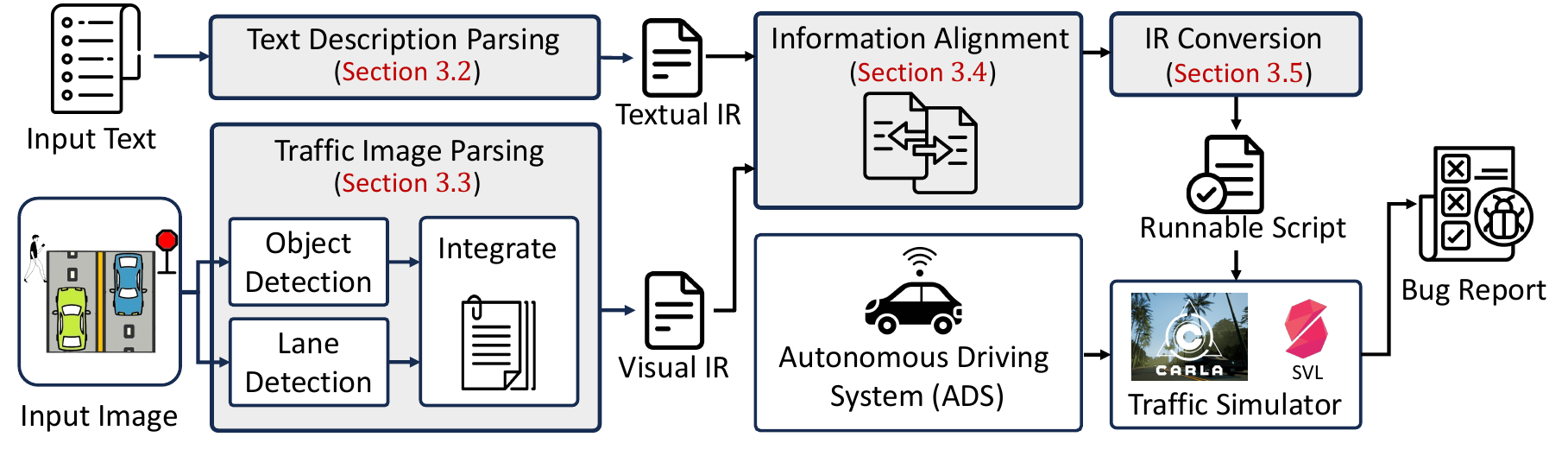}
    \caption{Overview of {\tool}.}
    \label{fig:pipeline}
\end{figure}
\section{Approach}
In this work, we introduce {\tool}, a multi-modal approach to mitigate the manual effort of constructing traffic scenarios for ADS testing. Figure~\ref{fig:pipeline} shows the workflow of {\tool}.
{\tool} leverages an LLM-based {\textualextractor} (Section~\ref{sec:text_parser}) and a novel {\visualextractor} (Section~\ref{sec:visual_parser}) to distill information from the two input modalities.
Then, {\tool} aligns the information from both modalities and encapsulates it into a comprehensive traffic IR (Section~\ref{sec:app_align}).
Subsequently, {\tool} automatically converts this integrated IR into an executable traffic scenario in a driving simulator (Section~\ref{sec:convert_ir}).

\subsection{Traffic Scenario Representation}
\label{sec:dsl}
\input{tables/table_dsl}
To effectively represent information extracted from two 
%% different 
modalities, we adopt and extend a Domain Specific Language (DSL) proposed by Deng et al.~\cite{deng2023target}.
We chose this DSL over other DSLs such as OpenSCENARIO~\cite{openScenario} and Scenic~\cite{fremont2019scenic, queiroz19} \revisecr{because} this DSL has a more concise syntax grammar and \revisecr{requires} less information to specify a traffic scenario. 
\revisepara{This concise grammar enables more accurate and efficient content generation using LLMs, improving the accuracy of the traffic scenarios generated by our framework.}

The DSL from Deng et al.~includes three main components: Environment, Road Network, and Actors, as shown in Figure~\ref{fig:dsl}. 
Each of these components is further divided into subcomponents to capture the specific semantic elements of traffic scenarios.
\reviseminor{
\textbf{Environment} characterizes the time of day and the weather conditions. 
\textbf{Road network} specifies road type, lane configurations, and traffic signals.
\textbf{Actor} describes each participant in a traffic scenario, including vehicles and pedestrians. For each actor, the DSL specifies the type, position, lane occupancy, and behavior (e.g., going forward, turning left, etc.).}
% \begin{itemize}
%     \item \textbf{Environment} characterizes the time of day and the weather conditions. 
%     \item \textbf{Road network} specifies road type, lane configurations, and traffic signals.
%     \item \textbf{Actor} describes each participant in a traffic scenario, including vehicles and pedestrians. For each actor, the DSL specifies the type, position, lane occupancy, and behavior (e.g., going forward, turning left, etc.).
% \end{itemize}

We extend this DSL by incorporating a lane index element for each actor.
This allows for the description of an actor's position by both rough relative position to the ego vehicle and the precise lane occupancies.
For example, to describe a car ``on the left of the ego vehicle'', the DSL supports a relative position representation, such as ``\emph{reference\_point}: \texttt{ego\_vehicle}, \emph{relative\_position}: \texttt{left}''. 
Simultaneously, it allows for precise lane occupancy specification, like ``\emph{lane\_idx}: \texttt{1}'' for the described vehicle and ``\emph{lane\_idx}: \texttt{2}'' for the ego vehicle.
This feature is fundamental to the DSL's role in aligning multi-modal inputs of {\tool}, as detailed in Section~\ref{sec:app_align}.

\input{tables/table_prompts}
\subsection{Textual Description Parser}
\label{sec:text_parser}
From the input textual description,  {\tool} identifies three primary categories of attributes: (1) environment aspects, including weather conditions and the layout of road networks; (2) dynamic information about the vehicles and pedestrians, such as the speed, forthcoming maneuvers, driving directions; and (3) rough positions of the surrounding vehicles.

Recent advances in Large Language Models (LLMs) \revisecr{demonstrate} their remarkable abilities in interpreting and reasoning text across various domains. 
Thus, {\tool} leverages LLMs to automatically extract traffic scenario information from the textual description \revisecr{using an} DSL.
%% In our implementation, 
We use GPT-4o~\cite{openai2024gpt4o}  \reviseminor{(gpt-4o-2024-05-13)} as the default LLM for {\tool}.
We also experimented with other LLMs, including Llama-3.1~\cite{dubey2024llama3herdmodels}, Claude~\cite{claudesonnet}, Mistral~\cite{mistrallarge2}, and Titan~\cite{titan}, as detailed in Section~\ref{sec:setup_llm}.
We use the default hyperparameter settings for each LLM in experiments.

Since LLMs are sensitive to prompts, we meticulously \revisecr{design} our prompts by integrating several effective prompting strategies, including role-playing~\cite{shanahan2023roleplay}, 
step-by-step instructions, 
\revisecr{chain-of-thought}~\cite{wei2022chain}, and in-context few-shot learning~\cite{NEURIPS20201457c0d6}. 
Specifically, our prompt is structured into \revisecr{four} segments, as illustrated in Table~\ref{tab:prompt_knowledge_extraction}. 
\revisecr{First}, the LLM is assigned the role of a domain expert in autonomous driving.
\revisecr{Second, the task is introduced through step-by-step reasoning using the chain-of-thought strategy.}
\revisecr{Third}, an example traffic scenario represented in the DSL (Section~\ref{sec:dsl}) is included to introduce the model with the grammar of the DSL. 
The \revisecr{last} segment of the prompt presents two \revisecr{examples, each consisting} of a text description as input and the corresponding desired traffic scenario represented in the DSL format as output.
\revisecr{This} is intended to familiarize the LLM with both the \revisecr{input structure} and the \revisecr{output DSL representation}.
Details on the prompt \revisecr{designs} are available in our source code repository~\cite{my_artifacts_prompt}.

However, textual descriptions suffer from inherent ambiguity and lack of precision.
\revisepara{Manually specifying intricate details in textual descriptions, such as vehicle color, model, and} \revisecr{precise} positioning, is cumbersome
%% and laborious 
for ADS developers, often requiring 10 to 20 sentences.
%% to provide sufficient details.
\revisepara{These details are important for ADS testing, as evidenced by a well-known Tesla autopilot crash~\cite{bbc_self_driving_cars_2016}. On May 7, 2016, a Tesla Model S operating on Autopilot failed to detect a white trailer against a bright sky, resulting in a fatal collision in Florida.}
%% This incident highlights the importance of specifying detailed attributes, such as vehicle color, in ADS testing.
\revisecr{Hence, to efficiently convey these essential details,} we need an additional source, a reference image of the traffic scene, as detailed in the next section.
%% Section~\ref{sec:visual_parser}.

\subsection{Visual Information Extractor}
\label{sec:visual_parser}
From the input reference image, {\tool} extracts three categories of attributes: (1) actors in the image; (2) spatial positioning of the actors; and (3) environmental details, such as traffic light status and the presence of traffic signs.

To accomplish this, {\tool} uses two advanced pre-trained computer vision models for object and lane detection.
For object detection, {\tool} uses YOLOv10~\cite{wang2024yolov10realtimeendtoendobject}, a state-of-the-art model pre-trained on the COCO dataset~\cite{lin2015microsoft}.
This model can detect $11$ types of objects relevant to traffic scenario descriptions, including pedestrians, six types of vehicles (e.g., car, truck, bus, etc.), traffic lights, traffic signs, parking meters, and fire hydrants.
It achieves an average precision of $54.4\%$ on the COCO~\cite{lin2015microsoft} validation set. In our application for detecting traffic-related objects, it achieves {\accyolo} precision, representing state-of-the-art performance in object detection.
For lane detection, {\tool} uses CLRNet~\cite{DBLP:conf/cvpr/ZhengHLTY0022}, a leading model in monocular lane detection~\cite{huang2023anchor3dlane, kirchmeyer2023convolutional, guo2023visual}.
Given a traffic scene image, CLRNet generates fine-grained pixel-level annotations for detected lane dividing lines.
It achieves F1 scores of $80.47$ and $96.12$ on two large-scale lane detection benchmarks, CULane~\cite{DBLP:conf/aaai/PanSLWT18} and LLAMAS~\cite{9022318}, demonstrating \revisecr{strong performance} in lane detection.

By integrating the detected positions of vehicles or pedestrians with lane markings, {\tool} can determine the specific lane occupied by each actor.
Similar to the {\textualextractor}, the {\visualextractor} represents the extracted information in the DSL.

\subsection{Aligning IR From the Two Modalities}
\label{sec:app_align}
{\tool} integrates information from two modalities by aligning the textual IR with the visual IR to generate the final IR for traffic scenarios.
\revisepara{For each component in the DSL, if conflicts arise between the two IRs, {\tool} adopts information from the IR with higher predefined priority.
The priority is determined based on the expressiveness of the corresponding modality for that component, as detailed below.}
If there are elements not specified by either modality, {\tool} marks them as unspecified.
% in the IR. 
These elements are assigned predefined values when generating executable traffic simulations, as detailed in Section~\ref{sec:convert_ir}.

\textbf{Environment.}
Textual descriptions are effective \revisecr{in} conveying abstract environmental configurations, such as time of day, weather, and traffic signal information.
Therefore, {\tool} prioritizes the textual IR for the \textit{Environment} components.
Although it is possible to infer this information from images, a single word in the textual description, such as "rainy" or "daytime," provides a more direct and reliable specification.

\textbf{Road Network.}
{\tool} uses the \emph{road types}, \emph{traffic signs} and \emph{traffic light} from textual IR, while it adpots the detailed \emph{lane number} from visual IR.
Although traffic signal information can be extracted from reference images, we prioritize the textual description as it provides a direct and reliable specification.

\textbf{Actors.}
For the \textit{Actors} element, {\tool} uses both textual and visual IRs. 
The textual descriptions specify the type of actors and their dynamic properties, such as speed and behavior, while the visual input provides accurate positional details, such as lane occupation.
Unlike other elements, where one modality is prioritized, {\tool} aligns and merges actor information from both IRs.
Actors are categorized into two groups---those mentioned in both modalities and those mentioned in only one.
For actors present in both, {\tool} aligns them based on location, considering actors from both modalities at the same position to be the same entity.
It then merges information from both modalities. In case of conflicts, {\tool} prioritizes the textual IR for dynamic properties and the visual IR for positional details.
For actors mentioned in only one modality, {\tool} checks for potential conflicts, such as overlapping positions, before incorporating them into the final IR.
If any actor features are not specified by either modality, {\tool} marks these features as unspecified in the IR.
These features are assigned predefined values when converting the IR to executable traffic simulations, as detailed in Section~\ref{sec:convert_ir}.

\subsection{Converting IR to Executable Traffic Scenario in Simulation}
\label{sec:convert_ir}
\begin{figure}[htb]
\centering
    \includegraphics[width=0.9\linewidth]{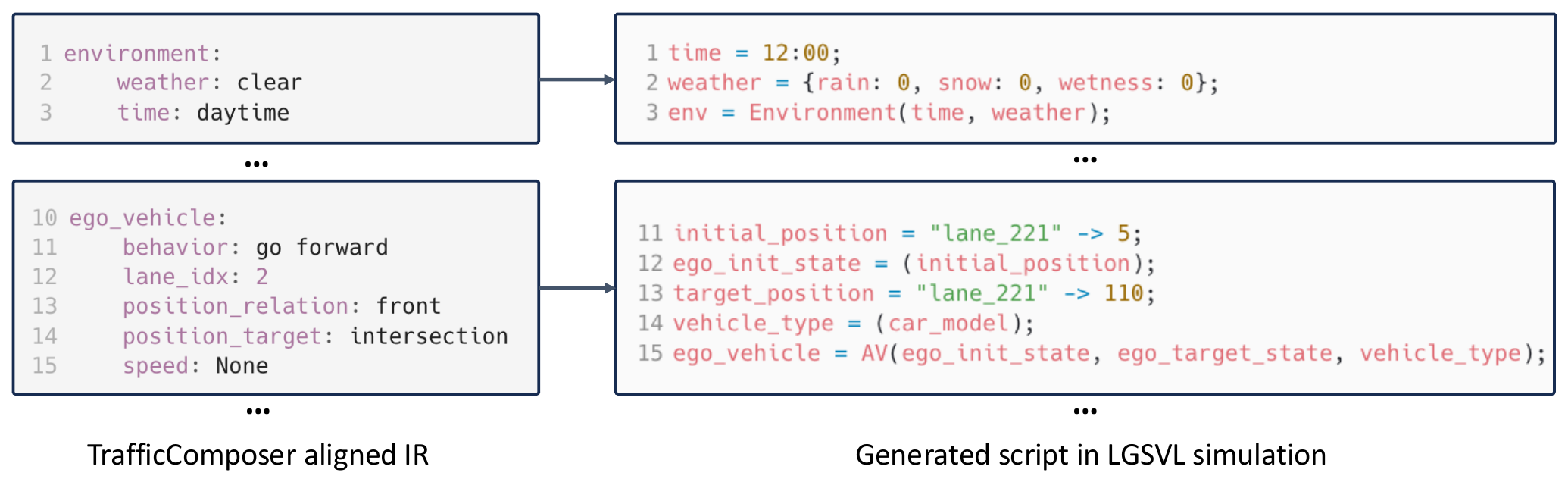}
\caption{Example of converting {\tool} generated IR to an executable script for LGSVL simulator.}
\label{fig:ir_convert}
\end{figure}

The aligned traffic IR provides a comprehensive description of the desired traffic scenario.
However, there is still a gap between this abstract traffic IR and executable traffic simulations.
Specifically, the traffic scenario generated by {\tool} is represented using an abstract DSL, as shown in Figure~\ref{fig:dsl}, which cannot be directly executed by a simulator.
To bridge the gap, {\tool} introduces an IR converter, which automates the generation of executable traffic simulations for each simulator.
In our experiments, we implement the IR converter for the two commonly used traffic simulators in recent ADS testing research, CARLA~\cite{DBLP:conf/corl/DosovitskiyRCLK17} and LGSVL~\cite{DBLP:conf/itsc/RongSTLLMBUGMAK20}. 
Figure~\ref{fig:ir_convert} illustrates a segment of the aligned IR and the corresponding generated script for the LGSVL simulator.

We abstract the process of generating traffic simulation scripts into three fundamental procedures: 
\texttt{set\_roadnetwork}, \texttt{set\_actor}, and \texttt{set\_environment}, reflecting the three core attributes of a traffic simulation.
{\tool} first searches for a map section that meets the required road network configurations, including \textit{road type}, \textit{lane number}, and \textit{traffic signals}. 
{\tool} then determines the initial position of the ego vehicle in the selected map section, using its \textit{lane\_idx} and relative position in the road network.
For instance, in the example shown in Figure~\ref{fig:ir_convert}, the ego vehicle is placed at \textit{``lane\_221''->5}.
The target position of the ego vehicle is then generated based on the specified \textit{behavior} in IR.
\revisecr{The positions of other actors} are calculated similarly, based on their \textit{lane\_idx} and their relative positions to the ego vehicle.
The environment configurations, including weather and time, are specified using the corresponding simulator APIs in the simulation script.
Since these operations are foundational \revisecr{for} traffic simulations, traffic simulators such as CARLA and LGSVL offer corresponding APIs to configure these elements~\cite{DBLP:conf/corl/DosovitskiyRCLK17, DBLP:conf/itsc/RongSTLLMBUGMAK20}.
For example, to add an actor to the traffic scenario, CARLA provides a function \texttt{carla.world.spawn\_actor()}, and LGSVL has the corresponding function \texttt{Simulator.add()}.

{\tool} also handles unspecified information in the aligned traffic IR.
For unspecified attributes in the \textit{Environment} component, such as weather conditions or time of day, {\tool} assigns default values from predefined candidate sets.
Information for the \textit{Road Network} component is reliably extracted from the reference image, ensuring no unspecified values.
For actors with unspecified positions, {\tool} assigns available locations while ensuring no overlap with existing actors in the scenario. 
For actors with unspecified dynamic attributes, default values are applied. 
{\tool} randomly samples a speed between 0 and 30 miles per hour and assigns straightforward driving behavior, maintaining realism in the simulation.

With the IR Converter, {\tool} detaches the generation of the traffic simulations from their design.
This separation grants {\tool} the flexibility to swiftly and effortlessly adapt to new traffic simulation platforms.
\todo{More examples are provided in the \textit{benchmark} folder of our research artifact~\cite{my_artifacts}}.

%% file: tables/table_dsl.tex
% \begin{figure}[htb]
\begin{wrapfigure}{r}{0.52\linewidth}
\centering
% \resizebox{0.6\linewidth}{!}{%
\resizebox{\linewidth}{!}{%
$\def\arraystretch{1.1}
\setlength{\arraycolsep}{1pt}
\begin{array}{lll}
\textbf{\emph{Scenario}} & ::= & \emph{Environment} ; \emph{Road\_network} ; \emph{Actors} \\
% \\
\textbf{\emph{Environment}} & ::= & \emph{weather}; \ \emph{time} \\
\emph{weather} & ::= & \texttt{rainy} \ | \ \texttt{foggy} \ | \ \texttt{snowy} \ | \ \texttt{wet} \ | \ \texttt{...} \\
\emph{time} & ::= & \texttt{daytime} \ | \ \texttt{nighttime} \\
% \\
\textbf{\emph{Road\_network}} & ::= & \emph{road\_type}; \ \emph{traffic\_signals}; \ \emph{lane\_number} \\
\emph{road\_type} & ::= & \texttt{intersection} \ | \ \texttt{roundabout} \ | \ \texttt{...} \\
\emph{traffic\_signals} & ::= & \emph{traffic\_signs}, \emph{traffic\_light} \\
\emph{traffic\_signs} & ::= & \epsilon \ | \ \emph{traffic\_sign}; \ \emph{traffic\_signs} \\
\emph{traffic\_sign} & ::= & \texttt{stop\_sign} \ | \ \texttt{speed\_limit\_sign} \ | \ \texttt{...} \\
\emph{traffic\_light} & ::= & \epsilon \ | \texttt{red\_light} \ | \ \texttt{green\_light} \\
\emph{lane\_number} & ::= & \texttt{0} \ | \ \texttt{1} \ | \ \texttt{2} \ | \ \texttt{3} \ | \ \texttt{...} \\
% \\
\textbf{\emph{Actors}} & ::= & \emph{ego\_vehicle}; \ \emph{npc\_actors} \\
\emph{ego\_vehicle} & ::= & \emph{behavior};  \ \emph{position} ; \ \emph{lane\_idx} \\
\emph{npc\_actors} & ::= & \epsilon \ | \ \emph{npc\_actor}; \ \emph{npc\_actors} \\
\emph{npc\_actor} & ::= & \emph{actor\_type}; \ \emph{behavior}; \ \emph{position} \\
\emph{actor\_type} & ::= & \texttt{car} \ | \ \texttt{truck} \ | \ \texttt{train} \ | \ \texttt{pedestrian} \ | \ \texttt{...} \\
\emph{behavior} & ::= & \texttt{go\_forward} \ | \ \texttt{turn\_left} \ | \ \texttt{static} \ | \ \texttt{...} \\
\emph{position} & ::= & \emph{reference\_point}; \ \emph{relative\_position} \\
\emph{reference\_point} & ::= & \texttt{ego\_vehicle} \ | \ \emph{road\_type} \ | \ \emph{traffic\_sign} \\
\emph{relative\_position} & ::= & \texttt{front} \ | \ \texttt{behind} \ | \ \texttt{left} \ | \ \texttt{on} \ | \ \texttt{...} \\
\emph{lane\_idx} & ::= & \texttt{0} \ | \ \texttt{1} \ | \ \texttt{2} \ | \ \texttt{3} \ | \ \texttt{...} \\
\end{array}$
}
\caption{The grammar of the traffic scenario DSL.}
\label{fig:dsl}
% \end{figure}
\end{wrapfigure}

%% file: tables/table_prompts.tex
\begin{table*}[thb]
    \centering
    \caption{The prompt for \reviseminor{textual description parser}.}
    \label{tab:prompt_knowledge_extraction}
\resizebox{0.9\textwidth}{!}{%
\begin{tabular}[c]{p{\textwidth}}
\hline
    % \rowcolor[HTML]{C0C0C0} \\[-7pt]
    \rowcolor[HTML]{C0C0C0} 
    \textbf{\normalsize Role setting}\\
    % \\[-5pt]
    % \hline
    \begin{tabular}[c]{@{}p{\textwidth}@{}}
        {\normalsize 
        You will work as an autonomous driving system testing engineer. Your task is to derive a test traffic scenario from the given traffic scenario description.
        
        }
    \\[-10pt]
    \end{tabular}\\
    % \\[-14pt]
\hline
    % \rowcolor[HTML]{C0C0C0} \\[-7pt]
    \rowcolor[HTML]{C0C0C0} 
    \textbf{\normalsize Breaks out of each step of the information parsing}\\
    % \\[-5pt]
% \hline
    \begin{tabular}[c]{@{}p{\textwidth}@{}}
    {\normalsize
        Approach this task step-by-step, take your time, and do not skip steps.
    \begin{itemize}[left=5.2ex]
        \item[Step 1:]Derive the `weather' element of the test scenario $\dots$
        \item[Step 2:]Derive the `time' of the test scenario $\dots$
        \item[Step 3:]Derive the test scenario's `road\_network' element. The `road\_network' element contains `road\_type', `traffic\_sign', `traffic\_light' and `lane\_number' $\dots$
        \item[Step 4:]Derive the `current\_behavior', `position\_target', `position\_relation', and `speed' for the ego vehicle $\dots$
        \item[Step 5:]Identify all actors other than the ego vehicle. For each actor, derive the following attributes: `type', `current\_behavior', `position\_target', `position\_relation', and `speed' $\dots$
        \item[Step 6:]Save all elements obtained above to a YAML, following the syntax below. Note that the final YAML output should start with `<YAML>' and end with `</YAML>'.
    \end{itemize}
    }
    \\[-10pt]
        % \{\textit{Details of DSL}\}
    \end{tabular} \\
\hline
    % \rowcolor[HTML]{C0C0C0} \\[-7pt]
    \rowcolor[HTML]{C0C0C0} 
    \textbf{\normalsize Grammar of the DSL}\\
    % \\[-5pt]
% \hline
    \begin{tabular}[c]{@{}p{\textwidth}@{}}
        \{\textit{\normalsize Details of DSL}\} \\
    \end{tabular}\\
    % \\[-5pt]

\hline
    % \rowcolor[HTML]{C0C0C0} \\[-7pt]
    \rowcolor[HTML]{C0C0C0} 
    \textbf{\normalsize Few-shot learning examples}\\
    % \\[-5pt]
    \begin{tabular}[c]{@{}p{\textwidth}@{}}
    {\normalsize
        Below is an example of an input traffic scenario description and the corresponding scenario representation with the DSL:
        
        Traffic description: \{\textit{content of the traffic description}\}
        
        Scenario representation: \{\textit{content of the scenario representation}\}
        
    }
        \\[-20pt]
        $$\cdots$$
        \\[-12pt]
    {\normalsize
        Based on the above descriptions and examples, convert the following traffic description to its corresponding scenario representation: 
        
        \{\textit{input traffic description}\} 
    }
    % \\[-5pt]
    \end{tabular} \\ 
\hline
\end{tabular}
}
\end{table*}

%% file: 3-Experiments.tex
\section{Experiment Design}
\label{sec:setup}
Our evaluation investigates six research questions about the effectiveness of {\tool}.

\begin{enumerate}
\item[\textbf{RQ1}] How closely are the traffic scenarios generated by {\tool} aligned with the original descriptions?
\item[\textbf{RQ2}] How effective are the traffic scenarios generated by {\tool} in identifying bugs in different ADSs?
\revisepara{\item[\textbf{\rqacceff}] How does scenario accuracy affect the effectiveness in exposing ADS failures?}
\item[\textbf{\rqablation}] To what extent does each modality contribute to the traffic scenario generation?
\item[\textbf{\rqrobustllm}] \revisepara{How sensitive is {\tool} to the choice of LLM and hallucination rates?}
\revisepara{\item[\textbf{\rqrobustcv}] How sensitive is {\tool} to the choice of computer vision models and object detection errors?}
\end{enumerate}

\subsection{Traffic Scenario Benchmark}
\label{sec:exp_benchmark}
To evaluate the effectiveness of {\tool}, we construct a \reviseminor{multi-model} benchmark comprising {\numscenario} traffic scenarios. 
\revisepara{Existing datasets such as nuScenes~\cite{caesar2020nuscenes} provide multi-modal sensor data, including camera images, radar signals, and Lidar signals from real-world vehicles. However, they do not include text descriptions or the counterpart tests in the simulation. 
In contrast, our benchmark provides textual descriptions, reference images, ground truth IRs, and corresponding simulations, enabling the evaluation of ADS test generation methods in simulation.}
These traffic scenarios are sourced from \revisepara{first-person perspective videos, including dashboard camera footages} from widely used datasets such as CULane~\cite{DBLP:conf/aaai/PanSLWT18}, and $11$ in-the-wild scenic drive videos on YouTube~\cite{youtubevideoNYC, youtubevideoFrankfurt, youtubevideoSeoul, youtubevideoShanghai}.
% youtubevideoSingapore
% youtubevideoDubai
% youtubevideoNewDelhi
% youtubevideoNYC2
% youtubevideoMiami
% youtubevideoSingaporeBeverly
% youtubevideoLA
To ensure diversity, we cluster the videos based on road network types and geographic locations and select representative frames as reference images of the scenarios.
\revisepara{The benchmark is designed to capture the diversity of and complexity of real-world traffic conditions, including low-light scenarios, day-night variations, significant occlusions, and both highway and urban roads across $9$ cities in $7$ countries.}
%% Please refer to our research artifact~\cite{my_artifacts} for more details.

\revisepara{For each scenario, a textual description is created following a structured guideline~\cite{benchmark_guideline}.
The textual description should include three core components: (1) environment, defined by time and weather conditions; (2) road network, including lane configurations and traffic control signals; and (3) actors, focusing on the two to three most relevant actors to the ego vehicle, with details including position, speed, and behavior.
The descriptions should be concise, as the reference images provide additional details.}
Ground truth traffic scenario IRs are annotated in the specialized DSL format.
\revisepara{With the textual description and the reference image combined, the ground truth IR of a traffic scenario is unique without ambiguity.}
Two authors independently annotated the ground truth IRs based on textual descriptions and reference images.
\revisepara{The initial annotations yield an average Tree Edit Distance (TED) of \result{$2.5$} and an average IE accuracy of \result{$97.7\%$}, as measured by the metrics detailed in Section~\ref{sec:setup_info_acc}.
A third author joins the reconciliation to resolve all disagreements.}
These ground truth traffic IRs serve as the basis for measuring the effectiveness of {\tool} in information extraction, as detailed in Section~\ref{sec:eval_acc}, Section~\ref{sec:eval_modality}, and Section~\ref{sec:eval_llms}.

\input{tables/tab_benchmarks}
\revisepara{\textit{Scenario richness evaluation.}
Compared to existing benchmarks in TARGET~\cite{deng2023target} and LawBreaker~\cite{DBLP:conf/kbse/0008P00Y22}, the proposed benchmark has more traffic scenarios with more complexity and higher diversity. As shown in Table~\ref{tab:benchmarks}, 
TARGET's scenarios are simple, each involving only one or two actors in addition to the ego vehicle, while LawBreaker includes only $10$ scenarios, with an average of $7$ actors per scenario.
In contrast, our proposed benchmark includes $120$ scenarios, each containing \revisecr{$5$ to $21$} actors, with an average of $7.8$ actors per scenario. 
To quantify dataset diversity, we compute the Vendi Score~\cite{DBLP:journals/tmlr/FriedmanD23}, a state-of-the-art metric for dataset diversity, following SCTrans~\cite{DBLP:conf/icse/DaiGLHLZY24}.
Our benchmark achieves a Vendi Score of \result{$22.1$}, significantly higher than TARGET's $17.0$ and LawBreaker's $3.1$.
These measurements underscore the benchmark's advantages in capturing diverse and complex traffic scenarios, making it suitable for evaluating advanced test generation methods.} % end of revise

\subsection{Experiment Setup for RQ1}
\label{sec:setup_info_acc}
To answer RQ1, we compare the generated traffic scenarios to the ground truth, using the benchmark described in Section~\ref{sec:exp_benchmark}.
\revisepara{We evaluate {\tool} against one single-modal LLM-based, six multi-modal LLM-based, and two traditional NLP method-based approaches, as detailed below.}

\reviseminor{The first baseline is} TARGET~\cite{deng2023target}, which uses the single-modal LLM GPT-4~\cite{openai2024gpt4technicalreport} to extract information from driving handbooks.
While TARGET is a strong baseline due to its use of LLM, it cannot process visual inputs.
Therefore, we provide only textual descriptions to TARGET in experiments.
We contacted TARGET's authors for the original implementation, \revisepara{and adopt their prompt design}.
Since the DSL used in {\tool} is extended on TARGET's DSL, to ensure a fair comparison, we exclude the extended components when calculating the accuracy for TARGET.

\revisepara{Furthermore, we consider six multi-modal LLMs, including GPT-4o~\cite{openai2024gpt4o} (\textit{gpt-4o-2024-05-13}), GPT-4o-mini~\cite{gpt-4o_mini} (\textit{gpt-4o-mini-2024-07-18}), Claude-3.5 Sonnet v2~\cite{claude3.5v2} (\textit{claude-3-5-sonnet-20241022-v2}), Claude-3.5 Sonnet v1~\cite{claudesonnet} (\textit{claude-3-5-sonnet-20240620-v1}), Llama 3.2 90B Vision Instruct~\cite{llama_3.2_90b_vision_instruct} (\textit{Llama-3.2-90B-Vision-Instruct}), and LLAVA~\cite{liu2023visual, llava1.6} (\textit{llava-v1.6-34b}).
Unlike TARGET, these models can process both the textual description and the reference image as inputs.
For these LLMs, we adapt the prompt design in {\tool} to handle textual inputs and further modify it to incorporate visual inputs. The adapted prompt integrates all the prompting strategies used in {\tool}, including role-playing, \revisecr{step-by-step instructions}, chain-of-thought reasoning, and in-context few-shot learning.}
Details on prompt designs are available in our source code repository~\cite{my_artifacts_prompt_mllm}.

\reviseminor{Finally,} we also consider two traffic scenario generation methods using traditional NLP techniques, AC3R~\cite{DBLP:conf/sigsoft/GambiHF19} and RMT~\cite{DBLP:journals/corr/abs-2012-10672}.
AC3R~\cite{DBLP:conf/sigsoft/GambiHF19} uses the Stanford Core NLP library~\cite{manning-etal-2014-stanford} and a domain-specific ontology to extract information about traffic crashes from police reports.
However, AC3R imposes strict requirements on input formatting, such as the need for detailed vehicle information (make, model, color, etc.), making it unable to process the free-form natural language descriptions of our traffic scenarios.
RMT~\cite{DBLP:journals/corr/abs-2012-10672} uses dependency parsing~\cite{kubler2009dependency} to extract information from natural language descriptions of traffic rules. 
RMT requires inputs to follow a strict ``if--then'' format, making it incapable of processing our free-form textual descriptions.

\textit{Evaluation metric.}
We compare a generated traffic scenario with the ground truth by measuring the tree edit distance (TED)~\cite{pawlik2016tree, pawlik2015efficient}. Tree edit distance is a commonly used metric to measure the similarity of programs in software engineering literature~\cite {4222572, 10.1145/1066157.1066243}.
\revisepara{In our experiments, TED quantifies both structure similarity and node similarity. When computing the distance between two IRs, missing or extra tree structures contribute to the distance based on the number of nodes involved. For tree structures that align, the algorithm further examines the values of individual nodes.}
\reviseminor{We calculate the information extraction (IE) accuracy based on the TED as follows:}
\begin{equation}
\label{eq:acc}
    \textit{IE Accuracy} (S; G) = 1 - \frac{TED(S, G)}{\textit{Total number of AST nodes (G)}}
\end{equation}
where $S$ is the generated traffic scenario in DSL format, and $G$ is the ground truth IR.
This accuracy metric measures how closely the generated scenario aligns with the ground truth.
\revisepara{To mitigate the impact of randomness, we repeat each experiment three times.}

\subsection{Experiment Setup for RQ2}
\label{sec:setup_fuzz}
To answer RQ2, we evaluate the bug detection capabilities of traffic scenarios generated by {\tool} in two ways.
First, we use these traffic scenarios as test cases to run ADS testing directly.
Second, we use these scenarios as initial seeds for ADS fuzzing methods. As shown by LawBreaker~\cite{DBLP:conf/kbse/0008P00Y22}, high-quality seeds can significantly boost the effectiveness of ADS fuzzing, which forms our hypothesis.

\textbf{Using generated traffic scenarios as direct test cases.}
We directly test {\numads} ADSs in the {\numscenario} traffic scenarios generated by {\tool} (detailed in Section~\ref{sec:exp_benchmark}).
These {\numads} ADSs are of different performance levels, ranging from industry-level ADSs to academic prototypes.
We include two Level 4~\cite{SAE2014Taxonomy} ADSs, Apollo~\cite{apollov6} and Autoware~\cite{8443742}, both of which are recognized for their robust performance and are frequently employed as benchmarks in ADS testing research~\cite{avfuzzer, DBLP:conf/kbse/0008P00Y22, DBLP:journals/tse/ZhongKR23, DBLP:conf/ccs/KimLRJ0K22, Feng2023Dense, DBLP:journals/tse/ZhangTTTSRTGWMNF23, DBLP:journals/tiv/ChenCZZ19}. %% Using Autoware
Additionally, we test on MMFN~\cite{9981775}, Roach~\cite{DBLP:conf/iccv/ZhangLDYG21}, and TransFuser~\cite{Chitta2023PAMI}---three cutting-edge ADSs proposed in top-tier conferences or journals, including IROS, CVPR, and PAMI.
In the end, we also included CARLA Behavior Agent~\cite{carlaBehaviorAgent}, a naive driving system in the CARLA simulator.
This ensures a comprehensive assessment covering a broad range of ADS capabilities, from advanced to basic ADSs.

\textbf{Using generated traffic scenarios as initial seeds in fuzzing methods.}
We test the {\numads} ADSs in \revisecr{fuzz testing} with two state-of-the-art ADS fuzzing methods, DriveFuzz~\cite{DBLP:conf/ccs/KimLRJ0K22} and LawBreaker~\cite{DBLP:conf/kbse/0008P00Y22}.

\textit{Details on the two fuzzing methods.}
DriveFuzz~\cite{DBLP:conf/ccs/KimLRJ0K22} tests whether the ADS encounters collisions, infractions, or immobility by mutating various factors of traffic scenarios.
These mutation operations include adding a new participant, altering the starting and ending points of a participant, modifying the location, size, and friction of puddles, and adjusting weather conditions.
LawBreaker~\cite{DBLP:conf/kbse/0008P00Y22} introduces a signal temporal logic-based DSL for specifying real-world traffic laws and a specification-coverage guided fuzzing algorithm for ADS testing.
It tests ADSs against traffic accidents and traffic law violations, covering 24 Chinese traffic laws.
LawBreaker supports mutations, including changes in position, speed, time, weather, and vehicle type.

\textit{Baselines.}
The effectiveness of the traffic scenarios generated by {\tool} is compared to two baseline seed settings: (1) using the original initial seeds of the fuzzing methods, and (2) using the traffic scenario generated by TARGET~\cite{deng2023target}.
We elaborate on each setting below.
\begin{enumerate}
    \item The initial seeds of DriveFuzz are randomly generated using predefined road maps in the CARLA simulator. 
Each scenario includes only one actor, with its position randomly assigned, and the weather condition is fixed to sunny.
DriveFuzz generates {\drivefuzzNumSeeds} initial seed traffic scenarios.
Before the fuzzing process, all seed scenarios are verified for legitimacy by dry-running the ADS under test and ensuring that it operates normally without encountering errors.
LawBreaker’s fuzzing process begins with ten high-quality traffic scenarios handcrafted by LawBreaker’s authors. These traffic scenarios are more complex than those in DriveFuzz, including, on average, \result{seven} participants other than the ego vehicle.
    \item Deng et al.~\cite{deng2023target} construct $98$ traffic scenarios using TARGET based on traffic laws in \revisecr{the} Texas Driver Handbook~\cite{texasdmv2022handbook}. Each scenario consists of the ego vehicle and one other actor.
\end{enumerate}
DriveFuzz is officially implemented on the CARLA simulator, while LawBreaker is on the LGSVL simulator.
Due to the discontinuation of the LGSVL simulator, the majority of our evaluated ADSs do not support LGSVL.
Thus, we can only run LawBreaker fuzzing experiments on LGSVL and Apollo.
\revisepara{LawBreaker relies on an outdated API, necessitating the use of Apollo v6.0 for compatibility.}

To ensure fair comparison, we randomly sample the same number of seeds as each baseline setting.
Specifically, for the comparison with DriveFuzz, we randomly sample {\drivefuzzNumSeeds} scenarios from the {\numscenario} scenarios generated by {\tool} in the benchmark.
For comparison with LawBreaker, we sample ten scenarios.
Before fuzzing, following the DriveFuzz~\cite{DBLP:conf/ccs/KimLRJ0K22} procedure, we pre-validate the sampled scenarios by dry-running the ADS under test to confirm they do not immediately cause crashes or traffic rule violations. 
If a scenario fails this validation, we resample and validate another from the benchmark.
For each experiment, we run the fuzzer for six hours, following the time budget setting from DriveFuzz~\cite{DBLP:conf/ccs/KimLRJ0K22}.
Table~\ref{tab:fuzzsetting} shows the experiment setting details.
\input{tables/tab_fuzz_setting}

\textit{Evaluation metrics.}
We evaluate both the effectiveness and efficiency of the fuzzing experiments.
We run fuzz testing on each ADS under \revisecr{the} three initial seed settings, within a six-hour time budget, following the setting in DriveFuzz~\cite{DBLP:conf/ccs/KimLRJ0K22}.
To assess bug detection \textit{effectiveness}, we measure the number of bugs detected in fuzzing experiments on each ADS.
For bug detection \textit{efficiency}, we measure the time taken to detect the first bug.
To mitigate the impact of randomness, each fuzzing experiment is repeated ten times.
To assess the statistical significance of the results, we conduct a one-way analysis of variance (ANOVA)~\cite{girden1992anova} on the data from these ten repetitions, followed by Tukey's HSD (honestly significant difference)~\cite{tukey1949comparing} post hoc analysis.

%%%% RQ: How does scenario generation accuracy affect the effectiveness in exposing ADS failures?
\subsection{\revisepara{Experiment Setup for {\rqacceff}}}
\label{sec:setup_acceff}
\revisepara{{\tool} and TARGET have significant differences in scenario generation accuracy, achieving $97.0\%$ and $74.3\%$ average IE accuracy, respectively (Section~\ref{sec:eval_acc}).
To answer {\rqacceff}, we use {\tool} and TARGET to generate tests based on the scenario descriptions in our benchmark and use them as initial seeds for ADS fuzzing methods.
To ensure a rigorous comparison, the fuzzing methods---DriveFuzz and LawBreaker---are configured to sample tests generated by {\tool} and TARGET from the same scenario as the seed each time, \revisecr{thus} isolating scenario accuracy as the only variable.}

\revisepara{\textit{Evaluation Metric.}
We measure the number of exposed ADS failures in a six-hour fuzzing on six ADSs, as in RQ2.
Each experiment is repeated ten times to mitigate the impact of randomness.}

\subsection{Experiment Setup for {\rqablation}}
\label{sec:setup_modality}
To answer {\rqablation}, we conduct an ablation study to assess the effectiveness of information extraction from each input modality. 
We create two variants of {\tool}---{\tooltext} processing only textual input and {\toolvisual} processing only visual input.
This evaluation follows the same experimental framework established in RQ1, ensuring consistency and comparability across different research questions.
We test {\tool} and its variants on the {\numscenario} traffic scenarios in our benchmark. 
Similar to RQ1, we use \revisecr{the} tree edit distance-based IE accuracy (Equation~\ref{eq:acc}) to measure the similarity of \revisecr{the} generated scenarios to the ground truth.
\revisepara{Each experiment is repeated three times to mitigate the impact of randomness.}

\subsection{Experiment Setup for {\rqrobustllm}}
\label{sec:setup_llm}
To answer {\rqrobustllm}, we evaluate the effectiveness of {\tool} when using different LLMs \revisepara{and when systematically introducing errors into the LLM-generated textual IRs}.

\reviseminor{\textit{Sensitivity to the choice of LLM}}. 
We construct {\numllm} variants of {\tool}, each using a different LLM.
By default, {\tool} employs GPT-4o~\cite{openai2024gpt4o} (\textit{gpt-4o-2024-05-13}) in its {\textualextractor}.
Additionally, we include GPT-4o-mini~\cite{gpt-4o_mini} (\textit{gpt-4o-mini-2024-07-18}) for comparison.
We also incorporate the open-source Llama-3.1~\cite{dubey2024llama3herdmodels} model in three sizes---\textit{Llama-3.1-8B-Instruct}, \textit{Llama-3.1-70B-Instruct}, and \textit{Llama-3.1-405B-Instruct}.
Furthermore, we use \result{three} additional closed-source LLMs: Claude~3.5~Sonnet~\cite{claudesonnet} \reviseminor{(\textit{claude-3-5-20240620})}, Mistral~Large~2~\cite{mistrallarge2} (\textit{mistral-large-2407}, with 123B parameters), and Amazon~Titan~Text~Premier~\cite{titan} (\textit{amazon.titan-text-premier-v1}).
These LLMs cover a wide range of parameter sizes, from $8$ billion to $405$ billion.

\textit{Sensitivity to LLM hallucination.}
\revisepara{To evaluate the sensitivity of {\tool} to LLM hallucinations, we conduct a controlled experiment by systematically injecting errors into the textual IR generated by LLM.
In the experiments, we use GPT-4o (\textit{gpt-4o-2024-05-13}), which generates textual IR with a $79.2\%$ IE accuracy (detailed in Section~\ref{sec:eval_modality}).
Errors are uniformly injected by randomly sampling a proportion of elements of the textual IR at a hallucination rate and mutating them to incorrect values. We experimented with 10 hallucination rates from $1\%$ to $10\%$ and quantified how the accuracy of {\tool} varies under different hallucination rates.}

\reviseminor{\textit{Evaluation Metrics.}} We evaluate the information extraction (IE) accuracy of traffic scenarios generated by these {\tool} variants by comparing them to the ground truth, using the same experimental settings as RQ1.
To ensure consistency and fairness, we apply \revisecr{an} identical prompt design \revisecr{in} all variants, as detailed in Section~\ref{sec:text_parser}.
All experiments are conducted using the default hyperparameter settings for each LLM.
To mitigate the impact of randomness, we repeat each experiment {\numrepeatllmexp} times and report the average results.

\subsection{\revisepara{Experiment Setup for {\rqrobustcv}}}
\label{sec:setup_robustcv}
\revisepara{
To answer {\rqrobustcv}, we evaluate the effectiveness of {\tool} when using different computer vision models and when systematically injecting errors into \revisecr{the} object detection results.}

\revisepara{
\textit{Sensitivity to the choice of object detection models.}
We construct {\numobjdet} variants of {\tool}, each using a different object detection model.
By default, {\tool} uses YOLOv10 (yolov10x) in {\visualextractor}.
Specifically, we include \result{four} earlier versions: YOLOv3 (yolov3u, released in March 2018), YOLOv5 (yolov5x6u, released in June 2020), YOLOv8 (yolov8x, released in January 2024), and YOLOv9 (yolov9e, released in February 2024).
In addition, we incorporate \result{five} size variants of YOLOv10: yolov10n, yolov10s, yolov10m, yolov10b, and yolov10l.}

\revisepara{
\textit{Sensitivity to object detection errors.}
To further assess {\tool}'s sensitivity to inaccuracies in object detection, we systematically inject errors into the detection results.
Based on the object detection results of yolov10x, which achieves a precision of $98.7\%$ on our benchmark images, errors are introduced by randomly removing detected actors at rates ranging from $1\%$ to $10\%$.}

\revisepara{
\textit{Evaluation Metric.}
We measure the information extraction (IE) accuracy of traffic scenarios generated by these {\tool} variants using the same settings as RQ1.
All experiments are conducted using the default hyperparameter settings for each object detection model.
Each experiment is repeated three times to reduce randomness.}

\section{Results}
\label{sec:eval}
\subsection{RQ1: Information Extraction (IE) Accuracy}
\label{sec:eval_acc}

\input{tables/tab_info_extract_baselines}
Table~\ref{tab:info_extract_baselines} shows the \reviseminor{average information extraction (IE)} accuracy of the generated traffic scenarios compared to the ground truth, as measured by Equation~\ref{eq:acc}.
\revisepara{We run each experiment three times and report the mean value and margin of error.}
{\tool} achieves a notable \reviseminor{average IE} accuracy of \result{$97.0\%$}, representing a significant absolute improvement of \result{$22.7\%$} over the single-modal LLM-based baseline TARGET, which achieves \result{$74.3\%$}.
\revisepara{
Among multi-modal LLM baselines, GPT-4o demonstrates the highest performance, achieving an average of \result{$89.7\%$} IE accuracy.}
However, {\tool} surpasses GPT-4o with an absolute improvement of \result{$7.3\%$}.
An in-depth analysis reveals further differences between {\tool} and GPT-4o.
Both methods perform well in extracting \textit{environment} configurations, such as weather and time of day, and the \textit{road network} configurations, e.g., road type and traffic signals.
The primary difference lies in the \textit{actors} element.
In \result{$41$} of the {\numscenario} traffic scenarios, GPT-4o fails to detect at least one vehicle or pedestrian, while {\tool} only makes such omissions in \result{$8$} cases.
% , outperforming GPT-4o by \result{$78.5\%$}. 
This highlights {\tool}'s superior effectiveness in actor-related information extraction, contributing to its overall \result{$7.3\%$} accuracy advantage over GPT-4o.

TARGET \reviseminor{cannot} extract information from visual inputs, leading to a significant performance gap compared to \reviseminor{multi-modal LLMs} and {\tool}.
As a multi-modal LLM, GPT-4o suffers from hallucination issues when processing visual inputs, such as generating configurations for a pedestrian that does not exist.
\revisepara{
{\tool} uses LLM in its {\textualextractor}. Therefore, it also suffers from hallucination problems. 
To evaluate the impact of LLM hallucinations in {\tool}, we randomly sample $24$ traffic scenarios ($20\%$ of our benchmark) and analyze the results of {\textualextractor}. 
{\tool} makes a total of \result{$4$} errors, including misidentified behaviors for \result{$2$} actors and incorrect speed assignment for \result{$2$} actors with no speed specification in the input textual description.} % end of revise
Compared to these baselines, {\tool} uses a more specialized and robust information extraction pipeline instead of solely relying on LLMs.
It leverages two pre-trained computer vision models to accurately extract the actors and lanes from the visual input.
A deterministic algorithm is then applied to analyze the information and generate the visual IR.
By aligning the information from both visual and textual inputs, {\tool} effectively mitigates the impact of hallucination on its overall performance.

\input{tables/tab_direct_bug}
\subsection{RQ2: Bug Detection Effectiveness and Efficiency}
\label{sec:eval_fuzz}
\textbf{Direct testing results.} 
Table~\ref{tab:direct_bug} shows the number of crashes or traffic rule violations detected by directly testing the {\numads} ADSs on the {\numscenario} traffic scenarios generated by {\tool}.
Overall, {\numdirectbug} of these {\numscenario} scenarios directly triggered crashes or traffic rule violations.
In particular, direct testing on the CARLA Behavior Agent triggers \result{$27$} crashes or violations.
Even on Apollo~\cite{apollov6}, the most robust ADS under test, \result{$7$} crashes are discovered.
These findings highlight the strong bug detection capabilities of the traffic scenarios generated by {\tool} in direct ADS testing.

\input{tables/fig_bug}
\input{tables/tab_bug}

\textbf{Fuzzing results.}
Figure~\ref{fig:bug} presents the number of crashes and traffic rule violations detected in fuzz testing over time under the three initial seed settings.
The lines represent average results from ten repeated experiment runs, and the shaded regions indicate variability across these ten runs.
Table~\ref{tab:bug} presents the average total number of crashes and violations detected in fuzz testing.
Across all {\numads} ADSs tested, scenarios generated by {\tool} detect, on average, \result{$53\%$--$117\%$} more bugs compared to the original fuzzing seeds and \result{$33\%$--$124\%$} more bugs than TARGET-generated scenarios.
Specifically, {\tool} detects on average \result{$77\%$} more bugs on Apollo and  \result{$99\%$} more on Autoware, compared to the best-performing baseline.
We conduct an
% a one-way 
analysis of variance (ANOVA) followed by Tukey’s HSD (honestly significant difference) \revisecr{post hoc} analysis.
The results of Tukey's HSD show that, across all {\numads} ADSs, the p-values for comparisons between {\tool} and the baseline settings range from \result{$0.0003$} to \result{$0.0062$}, all within the significance threshold of \result{$0.05$}.
This proves the statistical significance of {\tool}'s advancement over the two baselines.
Overall, these results demonstrate the effectiveness of {\tool}-generated scenarios as initial seeds for ADS fuzz testing.

\input{tables/tab_time_to_bug}
Furthermore, Table~\ref{tab:bug_time} shows the average time taken to detect the first crash or traffic rule violation across the ten runs.
Across all {\numads} ADSs, {\tool}-generated scenarios require on average \result{$28\%$--$58\%$} less time to find the first \revisecr{failure} compared to the original fuzzing seeds and \result{$30\%$--$52\%$} less time compared to TARGET.
The results of Tukey's HSD on the time consumption show that, across all {\numads} ADSs, the p-values for comparisons between {\tool} and the baseline settings range from \result{$0.0018$} to \result{$0.0053$}, all within the significance threshold of \result{$0.05$}.
This proves the scenarios generated by {\tool} can significantly improve the time efficiency of ADS fuzz testing as initial seeds.

\input{tables/tab_bug_ratio}
\revisepara{\textit{Manual analysis of exposed ADS failures.}
To assess whether the exposed crashes and traffic rule violations are attributable to the ADSs under test, we manually review all exposed failures across six ADSs. 
\textbf{In the direct testing experiments}, all crashes and violations are caused by ADS errors. \textbf{In the fuzzing experiments}, at least \result{$80.1\%$} of the exposed failures are due to ADS errors among the six ADSs. 
Table~\ref{tab:bug_ratio} summarizes the proportion of failures attributed to each ADS in the ten repeated fuzzing experiments.
Specifically, for the two industrial-level ADSs, \result{$82.9\%$} of \result{$391$} failures in Apollo and $86.5\%$ of $416$ failures in Autoware are due to ADS errors. 
Among the remaining $13.5\%$ of failures not caused by Autoware, \result{$9.1\%$} result from abnormal behavior of other actors. For example, vehicles unexpectedly collided with the ego vehicle from the rear. The other \result{$4.4\%$} is due to the abnormal spawn point of other actors, for example, pedestrians suddenly appearing on highways.
These issues, introduced by random mutations in the fuzzing process, could potentially be mitigated by refining the mutation algorithms and introducing additional constraints.} % end of revise
Figure~\ref{fig:case} showcases traffic scenarios where ADSs crash or violate traffic rules with more details.
Specifically, Figure~\ref{fig:case_apollo} depicts a scenario in which Apollo \revisecr{runs} a red light at an intersection, creating a hazardous situation with an oncoming left-turning vehicle.
Figure~\ref{fig:case_autoware} shows a scenario where Autoware \revisecr{fails} to yield when merging into the left lane, resulting in a collision with a vehicle in the target lane that has the right of way.
\revisepara{These results further prove the effectiveness of {\tool} generated traffic scenarios in exposing ADS failures.}

\input{tables/fig_case}

Across ADSs of varying performance levels, direct testing on the traffic scenarios generated by {\tool} successfully discovers \revisecr{at least}  seven crashes or traffic rule violations.
When these scenarios are used as initial seeds in fuzz testing, significant improvements in both bug detection effectiveness and time efficiency are observed.
Additionally, the detailed crash and traffic rule violations \revisecr{cases presented} in Figure~\ref{fig:case} further highlight the effectiveness of these scenarios. 
\revisepara{The analysis on Table~\ref{tab:bug_ratio} proves the effectiveness of {\tool} generated traffic scenarios in identifying ADS failures.}
Collectively, the results demonstrate that the traffic scenarios generated by {\tool} are highly effective for detecting crashes and traffic rule violations.

\input{tables/tab_acceff}

\subsection{\revisepara{{\rqacceff}: Bug Detection Effectiveness on Different Scenario Accuracy}}
\revisepara{Table~\ref{tab:acceff} shows the average number of ADS failures exposed in fuzzing experiments using the same seed scenarios with different accuracy (\result{$74.3\%$} by TARGET and \result{$97.0\%$} by {\tool}, respectively).
Across all {\numads} ADSs, scenarios with \result{$97.0\%$} accuracy expose \result{$33\%$--$109\%$} more failures compared to scenarios with \result{$74.3\%$} accuracy. 
Specifically, on the two industrial-level ADSs, the high-accuracy scenarios detected an average of \result{$62\%$} more bugs in Apollo and \result{$76\%$} more in Autoware.
These results demonstrate that scenarios with higher accuracy are more effective in exposing failures than \revisecr{those} with lower accuracy.} % end of revise

\subsection{{\rqablation}: Contribution of Each Modality}
\label{sec:eval_modality}
\input{tables/tab_info_extract_ablation}
Table~\ref{tab:ablation} shows the contribution of textual and visual inputs toward the \reviseminor{information extraction (IE)} accuracy of the generated traffic scenarios.
{\tool} achieves an \reviseminor{average IE} accuracy of \result{$97.0\%$}, demonstrating an absolute improvement of \result{$17.8\%$} and \result{$45.2\%$}, compared to the textual input-only and visual input-only variants.
\revisepara{Notably, {\tool}'s {\visualextractor} uses deterministic computer vision models that produce identical outputs for the same input.
Thus, IE accuracy exhibits a margin of error of $0$ when using visual input only.}
The results indicate that {\tool}'s integrated approach significantly enhances information extraction accuracy compared to relying on a single modality. 
Text inputs, though expressive, are limited by their inherent ambiguity, resulting in information loss.
While visual inputs provide precise spatial information, they lack the ability to convey dynamic aspects, such as vehicle speed or maneuver.
Aligning these two complementary modalities, {\tool} effectively mitigates these limitations, ensuring a comprehensive and accurate representation of traffic scenarios.

\subsection{{\rqrobustllm}: Sensitivity to the Choice of LLMs \revisepara{and Hallucination Rates}}
\label{sec:eval_llms}
\input{tables/tab_info_extract_llm}
\reviseminor{\textit{Sensitivity to the choice of LLMs.}}
Table~\ref{tab:llm} presents the information extraction (IE) accuracy of {\tool}, using {\numllm} LLMs from {\numllmfamily} LLM families.
% \todo{Due to space limit, we move some evaluation results to supplementary material~\cite{my_artifacts_supplementary}.}
We run each experiment {\numrepeatllmexp} times and report the mean value and margin of error.
%% in Table~\ref{tab:llm}.
Across all tested LLMs, {\tool} demonstrates robust performance, with average IE accuracy ranging from \result{$90.7\%$} to \result{$97.0\%$}.
Although there is a noticeable performance difference between the higher-capacity GPT-4o (\textit{gpt-4o-2024-05-13}) and the Titan (\textit{amazon.titan-text-premier-v1}), {\tool} continues to achieve acceptable results even with the least effective LLM.

\noindent \textit{Sensitivity to LLM hallucination.} \revisepara{Table~\ref{tab:error_llm} shows the information extraction (IE) accuracy of {\tool} when hallucinations are systematically injected into the textual IR generated by GPT-4o. The base text parsing accuracy of GPT-4o is $79.2\%$. With the complement from the visual modality, {\tool} is able to maintain an accuracy above \result{$90\%$} when up to $5\%$ hallucinations are injected into the original output of GPT-4o. Overall, there is a negative linear correlation between the LLM hallucination rate and {\tool}'s IE accuracy.} % end of revise

\subsection{\revisepara{{\rqrobustcv}: Sensitivity to the Choice of CV Models and Object Detection Errors}}
\input{tables/tab_info_extract_det+error_det}
\revisepara{\textit{Sensitivity to the choice of CV models.}
Table~\ref{tab:objdet} shows the average information extraction (IE) accuracy of {\tool} using ten object detection models released from 2018 to 2024 with model sizes ranging from 2.3 million to 155.4 million parameters.
From the older yolov3u model released in 2018 to the more powerful yolov10x (the default model in {\tool}), {\tool} consistently achieves high average IE accuracy ranging from \result{$94.1\%$ to $97.0\%$}.}

\noindent \textit{Sensitivity to object detection errors.}
\revisepara{
Table~\ref{tab:error_det} shows the IE accuracy of {\tool} under injected object detection errors and the corresponding object detection precision on our benchmark reference images.
The base detection precision of yolov10x is $98.7\%$.
When error rates reach \result{$3\%$}, {\tool} maintains an average IE accuracy above \result{$90\%$}.
The results demonstrate a negative linear correlation between the detection error rate and {\tool}'s IE accuracy.} % end of revise

%% file: tables/tab_benchmarks.tex
\begin{wraptable}{r}{0.5\textwidth}
% \begin{table}[]
\centering
\caption{\revisepara{Comparison of three traffic scenario benchmarks on scale, complexity, and diversity.}}
\label{tab:benchmarks}
\resizebox{\linewidth}{!}{%

\revisepara{
\begin{tabular}{lccc}
\toprule
Benchmark      & \# Scenarios         & \# Actors       & Vendi Score~\cite{DBLP:journals/tmlr/FriedmanD23} \\ \midrule
LawBreaker     & \phantom{0}10      & \phantom{0.}7      & \phantom{0}3.1        \\
TARGET         & \phantom{0}98      & \phantom{0.}1      & 17.0       \\
\textbf{\tool} & \textbf{120}       & \textbf{7.8}       & \textbf{22.1}       \\ 
\bottomrule
\end{tabular}%
} % end of revise

} % end of resizebox
% \end{table}
\end{wraptable}

%% file: tables/tab_fuzz_setting.tex
% \begin{wraptable}{r}{0.65\textwidth}
\begin{table}[htb]
\centering
\caption{Experiment setup for the ADS fuzzing experiments under three initial seed settings.}
\label{tab:fuzzsetting}
\resizebox{0.8\linewidth}{!}{%

\renewcommand{\arraystretch}{0.33}  % Reduce row height
\begin{tabular}{c|c|l|l|c|c}
% \toprule
% \hline
\specialrule{1pt}{0pt}{0pt}  % Same thickness as \toprule
\multirow{5}{*}{Fuzzing Method} & \multirow{5}{*}{Simulator} & \multirow{5}{*}{ADS under test} & \multirow{5}{*}{Setting} & \multirow{5}{*}{\# of seeds} & \multirow{5}{*}{Time (h)} \\ 
 &  &  &  &  & \\
  &  &  &  &  & \\
   &  &  &  &  & \\
    &  &  &  &  & \\
\hline
\multirow{15}{*}{LawBreaker} & \multirow{15}{*}{LGSVL} & \multirow{15}{*}{Apollo~\cite{apollov6}} & \multirow{5}{*}{original} & \multirow{5}{*}{10} & \multirow{5}{*}{6} \\
 &  &  &  &  & \\ 
  &  &  &  &  & \\ 
   &  &  &  &  & \\ 
    &  &  &  &  & \\ 
\cline{4-6}
% -----
 &  &  & \multirow{5}{*}{TARGET~\cite{deng2023target}} & \multirow{5}{*}{10} & \multirow{5}{*}{6} \\ 
 &  &  &  &  & \\ 
  &  &  &  &  & \\ 
   &  &  &  &  & \\ 
    &  &  &  &  & \\ 
\cline{4-6}
% -----
 &  &  & \multirow{5}{*}{{\tool}} & \multirow{5}{*}{10} & \multirow{5}{*}{6} \\ 
 &  &  &  &  & \\ 
  &  &  &  &  & \\ 
   &  &  &  &  & \\ 
    &  &  &  &  & \\ 
\hline
% 1
\multirow{15}{*}{DriveFuzz} & \multirow{15}{*}{CARLA} & 
% \multirow{15}{*}{
% \renewcommand{\arraystretch}{0.7}  % Reduce row height
% \begin{tabular}[c]{@{}l@{}}Autoware\\ MMFN\\ TransFuser\\ Roach\\ Behavior Agent\end{tabular}
% } 
\multirow{3}{*}{Autoware~\cite{8443742}}
& \multirow{5}{*}{original} & \multirow{5}{*}{40} & \multirow{5}{*}{6} \\
% 2
 &  &  &  &  & \\ 
% 3
  &  &  &  &  & \\ 
\cline{3-3}
% 4
& & \multirow{3}{*}{MMFN~\cite{9981775}} & & & \\
% 5
 &  &  &  &  & \\ 
\cline{4-6}
% 6
  &  &  & \multirow{5}{*}{TARGET~\cite{deng2023target}} & \multirow{5}{*}{40} & \multirow{5}{*}{6} \\ 
\cline{3-3}
% 7
& & \multirow{3}{*}{TransFuser~\cite{Chitta2023PAMI}} & & & \\
% 8
 &  &  &  &  & \\ 
% 9
  &  &  &  &  & \\ 
\cline{3-3}
% 10
& & \multirow{3}{*}{Roach~\cite{DBLP:conf/iccv/ZhangLDYG21}} & & & \\
\cline{4-6}
% 11
 &  &  & \multirow{5}{*}{\tool} & \multirow{5}{*}{40} & \multirow{5}{*}{6} \\ 
% 12
  &  &  &  &  & \\ 
\cline{3-3}
% 13
& & \multirow{3}{*}{Behavior Agent~\cite{carlaBehaviorAgent}} & & & \\
% 14
 &  &  &  &  & \\ 
% 15
  &  &  &  &  & \\ 
% \bottomrule
% \hline
\specialrule{1pt}{0pt}{0pt}  % Same thickness as \toprule
\end{tabular}%

}
% \end{wraptable}
\end{table}

%% file: tables/tab_info_extract_baselines.tex
\begin{wraptable}{r}{0.39\textwidth}
\centering

\caption{Comparison of average IE accuracy using different methods.}
\label{tab:info_extract_baselines}
\resizebox{\linewidth}{!}{%

\begin{tabular}{lc}
\toprule
Method & IE Acc. ($\%$) \\ \midrule
TARGET & $74.3$\revisepara{$(\pm 1.5)$} \\
gpt-4o-2024-05-13 & $89.7$\revisepara{$(\pm 5.2)$} \\
\revisepara{gpt-4o-mini-2024-07-18} & \revisepara{\result{$79.1(\pm6.1)$}}     \\
\revisepara{claude-3-5-sonnet-20241022-v2}   & \revisepara{$86.1(\pm4.3)$}      \\
\revisepara{claude-3-5-sonnet-20240620-v1}   & \revisepara{$84.1(\pm4.9)$}     \\
\revisepara{Llama-3.2-90B-Vision-Instruct} & \revisepara{$80.2(\pm6.1)$}      \\
\revisepara{llava-v1.6-34b}                         & \revisepara{$72.2(\pm4.2)$} \\
\textbf{\tool} & \textbf{97.0}\revisepara{\textbf{$(\pm 1.2)$}} \\ \bottomrule
\end{tabular}%

}

\end{wraptable}

%% file: tables/tab_direct_bug.tex
% \begin{wraptable}{r}{0.7\textwidth}
\begin{table}[htb]
\centering
\caption{Number of crashes or traffic rule violations detected in direct testing ADSs on {\tool}-generated traffic scenarios.}
\label{tab:direct_bug}
\resizebox{0.65\linewidth}{!}{%
\begin{tabular}{lllllll}
\toprule
 & Apollo & Autoware & MMFN & TransFuser & Roach & Behavior Agent \\ \midrule
\# \revisecr{failures} & \multicolumn{1}{c}{7} & \multicolumn{1}{c}{9} & \multicolumn{1}{c}{19} & \multicolumn{1}{c}{24} & \multicolumn{1}{c}{14} & \multicolumn{1}{c}{27} \\ \bottomrule
\end{tabular}%
}
\end{table}

%% file: tables/fig_bug.tex
\begin{figure}[th]
    \begin{center}
    % First row with
    \begin{subfigure}[b]{0.32\linewidth}
        \centering
        \includegraphics[width=\linewidth]{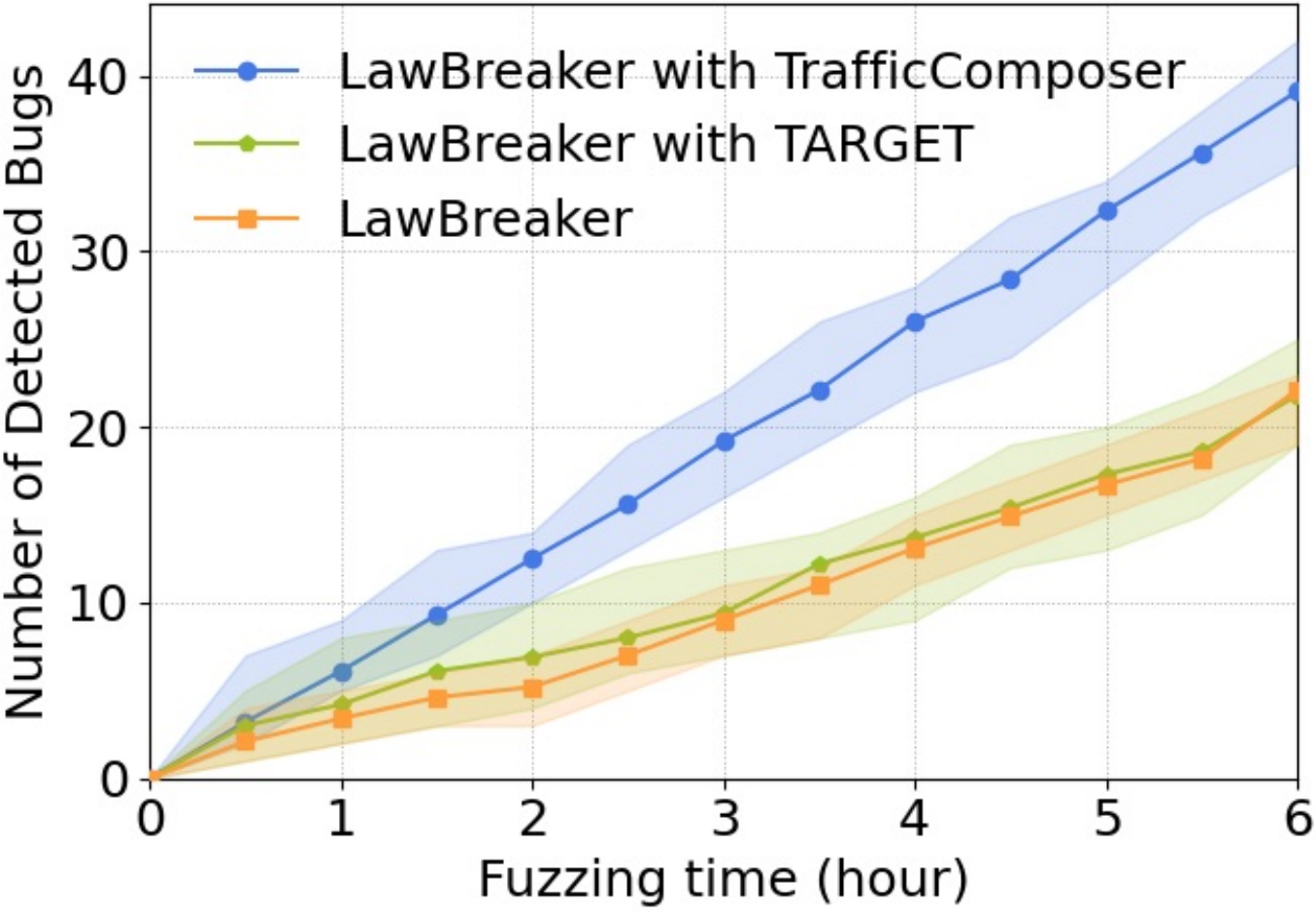}
        \caption{Apollo~\cite{apollov6}}
        \label{fig:bug_apollo}
    \end{subfigure}
    \hfill % spacing between the subfigures
    \begin{subfigure}[b]{0.32\linewidth}
        \centering
        \includegraphics[width=\linewidth]{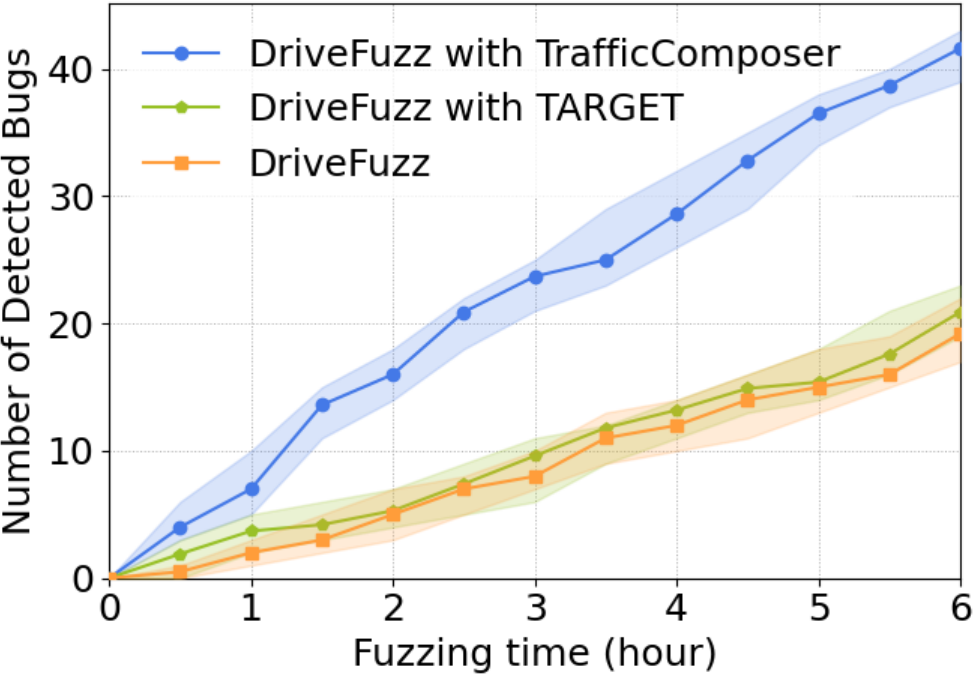}
        \caption{Autoware~\cite{8443742}}
        \label{fig:bug_autoware}
    \end{subfigure}
    \hfill % spacing between the subfigures
    \begin{subfigure}[b]{0.32\linewidth}
        \centering
        \includegraphics[width=\linewidth]{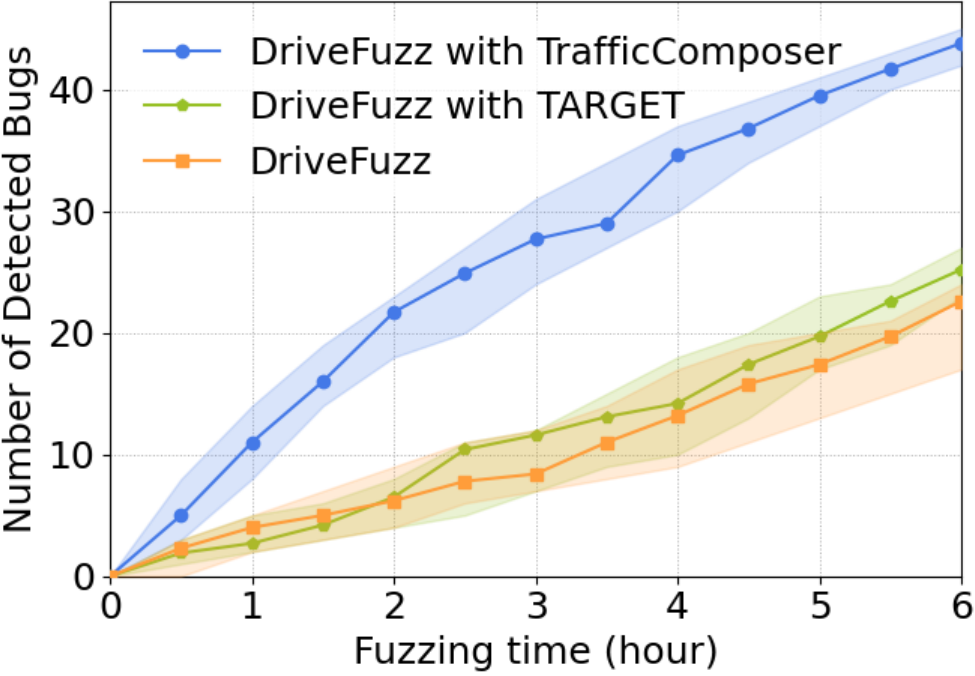}
        \caption{MMFN~\cite{9981775}}
        \label{fig:bug_mmfn}
    \end{subfigure}
    \end{center}

    % Second row
    \begin{subfigure}[b]{0.32\linewidth}
        \centering
        \includegraphics[width=\linewidth]{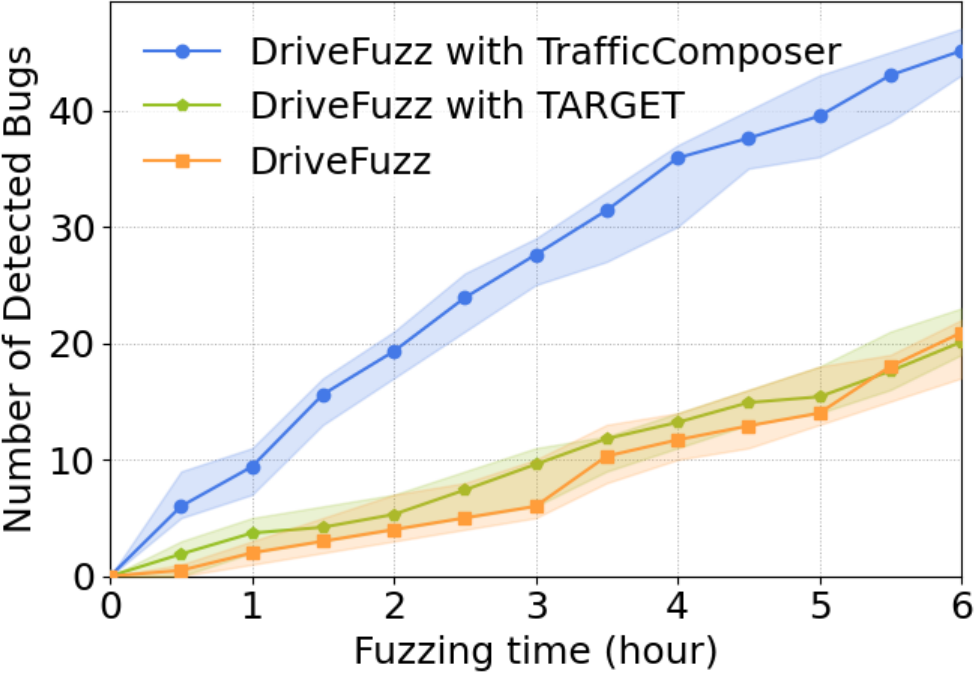}
        \caption{TransFuser~\cite{Chitta2023PAMI}}
        \label{fig:bug_transfuser}
    \end{subfigure}
    \hfill
    \begin{subfigure}[b]{0.32\linewidth}
        \centering
        \includegraphics[width=\linewidth]{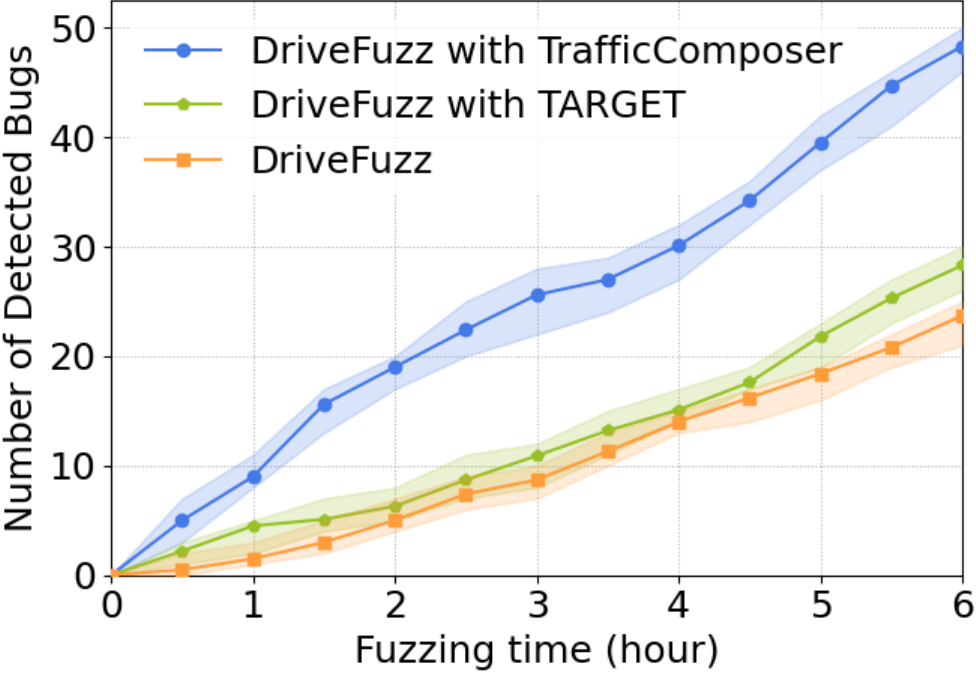}
        \caption{Roach~\cite{DBLP:conf/iccv/ZhangLDYG21}}
        \label{fig:bug_roach}
    \end{subfigure}
    \hfill
    \begin{subfigure}[b]{0.32\linewidth}
        \centering
        \includegraphics[width=\linewidth]{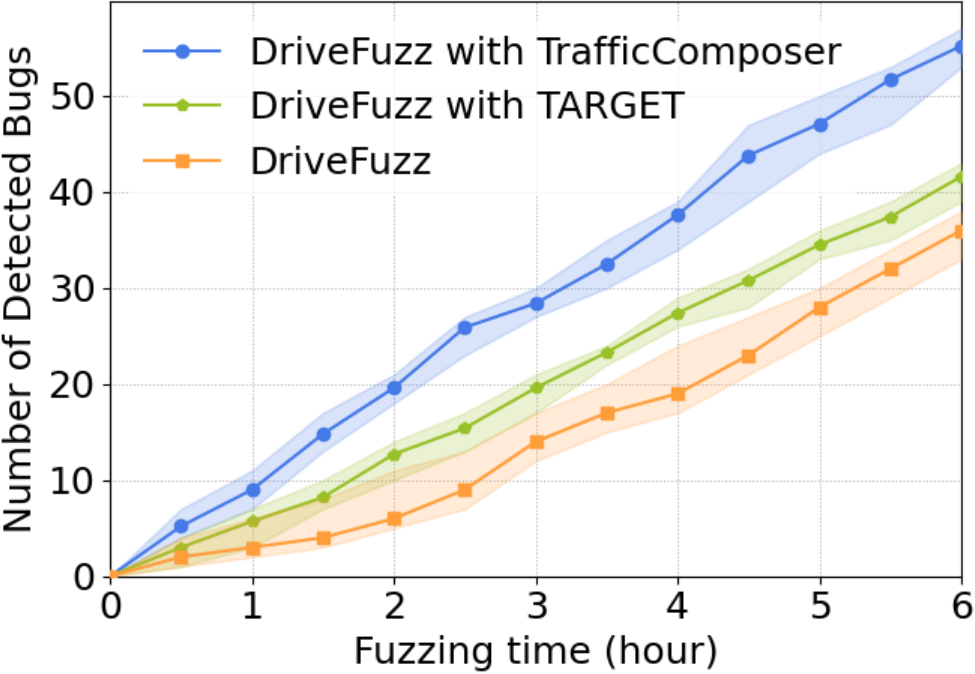}
        \caption{Behavior Agent~\cite{carlaBehaviorAgent}}
        \label{fig:bug_carla_behavior}
    \end{subfigure}

    \caption{Number of detected crashes and violations over testing time on {\numads} ADSs. The lines represent the average results of ten repeated experiments. The shaded regions indicate the variability across the ten runs.}
    \label{fig:bug}
\end{figure}

%% file: tables/tab_bug.tex
\begin{table}[htb]
\centering
\caption{Average number of detected crashes and violations in fuzz testing on the ADSs under test.}
\label{tab:bug}
\resizebox{0.85\linewidth}{!}{%

% \begin{tabular}{lcccccc}
% \toprule
% Model & Apollo & Autoware & MMFN & TransFuser & Roach & Behavior Agent \\ \hline
% DriveFuzz & - & 19.2 & 24.1 & 21.9 & 24.7 & 38.7 \\
% LawBreaker & 22.1 & - & - & - & - & - \\
% TARGET & 19.2 & 20.1 & 25.3 & 19.6 & 28.2 & 40.2 \\
% \textbf{\tool} & 43.3 & 46.2 & 48.1 & 50.0 & 52.1 & 55.2 \\ 
% \bottomrule
% \end{tabular}%

\begin{tabular}{l|c|ccccc}
% \toprule
\specialrule{1pt}{0pt}{0pt}  % Same thickness as \toprule
 & \multicolumn{1}{l|}{LawBreaker} & \multicolumn{5}{c}{DriveFuzz} \\ 
\cline{2-7}
Initial Seed Setting & Apollo & Autoware & MMFN & TransFuser & Roach & Behavior Agent \\ 
% \midrule
\hline
original & 22.1 & 19.2 & 22.6 & 20.9 & 23.7 & 36.0 \\
TARGET & 21.8 & 20.9 & 25.2 & 20.1 & 28.3 & 41.6 \\
\textbf{\tool} & \textbf{39.1} & \textbf{41.6} & \textbf{43.8} & \textbf{45.1} & \textbf{48.3} & \textbf{55.2} \\ 
% \bottomrule
\specialrule{1pt}{0pt}{0pt}  % Same thickness as \toprule
\end{tabular}%

}
\end{table}

%% file: tables/tab_time_to_bug.tex
\begin{table}[htb]
\centering
\caption{Average time consumed (in minutes) to detect the first crash or violation in ADS fuzz testing.}
\label{tab:bug_time}
\resizebox{0.85\linewidth}{!}{%

% \begin{tabular}{lccccccc@{}}
% \toprule
% Method & Apollo & Autoware & MMFN & TransFuser & Roach & Behavior Agent \\ \midrule
% DriveFuzz & - & 32 & 42 & 43 & 21 & 13 \\
% LawBreaker & 45 & - & - & - & - & - \\
% TARGET & \textbf{21} & \textbf{18} & \textbf{24} & \textbf{24} & \textbf{16} & \textbf{11} \\
% \textbf{{\tool}} & \textbf{21} & \textbf{18} & \textbf{24} & \textbf{24} & \textbf{16} & \textbf{11} \\
% \bottomrule
% \end{tabular}%

\begin{tabular}{l|c|ccccc}
% \toprule
\specialrule{1pt}{0pt}{0pt}  % Same thickness as \toprule
 & \multicolumn{1}{l|}{LawBreaker} & \multicolumn{5}{c}{DriveFuzz} \\ 
\cline{2-7}
Initial Seed Setting & Apollo & Autoware & MMFN & TransFuser & Roach & Behavior Agent \\ 
% \midrule
\hline
original & 31.6 & 21.4 & 25.2 & 22.7 & 18.4 & 12.7 \\
TARGET & 27.8 & 21.8 & 18.7 & 21.4 & 19.3 & 10.4 \\
\textbf{\tool} & \textbf{13.4} & \textbf{15.3} & \textbf{12.7} & \textbf{10.3} & \phantom{0}\textbf{9.7} & \phantom{0}\textbf{6.1} \\ 
% \bottomrule
\specialrule{1pt}{0pt}{0pt}  % Same thickness as \toprule
\end{tabular}%

}
\end{table}

%% file: tables/tab_bug_ratio.tex
\begin{table}[htb]
\centering
\caption{\revisepara{Failures detected in ten fuzzing experiments and proportion attributed to the ADS.}}
\label{tab:bug_ratio}
\resizebox{0.88\columnwidth}{!}{%

\revisepara{
\begin{tabular}{l|c|ccccc}
% \hline
\specialrule{1pt}{0pt}{0pt}  % Same thickness as \toprule
                                  & LawBreaker & \multicolumn{5}{c}{DriveFuzz}                         \\ \cline{2-7} 
                                  & Apollo     & Autoware & MMFN & TransFuser & Roach & Behavior Agent \\ 
\hline
total \# of bugs in 10 runs     & 391        & 416      & 438  & 451        & 483   & 552            \\ 
Proportion due to ADS (\%) &  82.9    & 86.5     & 80.1  & 83.0     & 88.0   &  81.0      \\
% \hline
\specialrule{1pt}{0pt}{0pt}  % Same thickness as \toprule
\end{tabular}%
} % end of revise

} % end of resizebox
\end{table}

%% file: tables/fig_case.tex
% \begin{wrapfigure}{r}{0.5\textwidth}
\begin{figure}[htb]
\centering
\resizebox{\linewidth}{!}{
\begin{minipage}{\linewidth}
    % \hfill
    \begin{subfigure}[b]{0.32\linewidth}
        \centering
        \includegraphics[width=\linewidth]{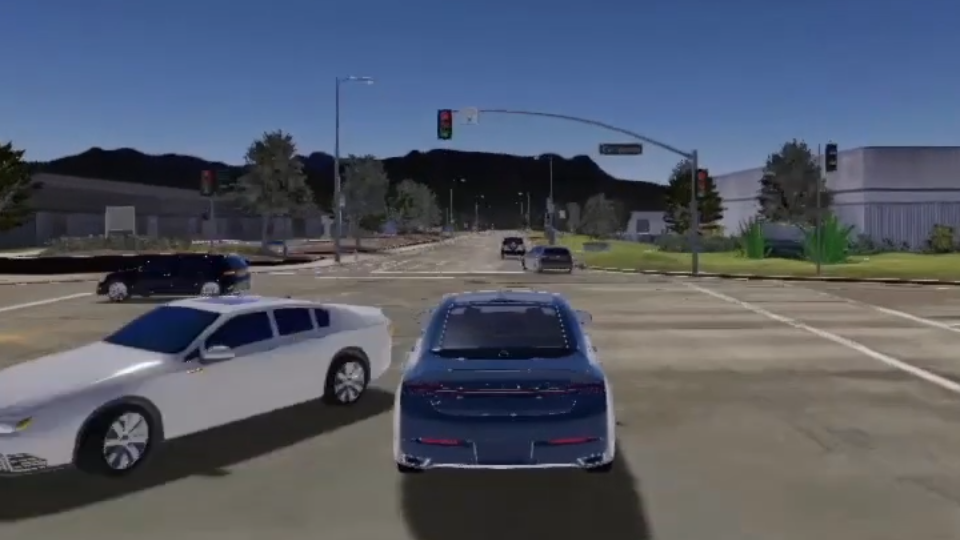}
        \caption{Apollo~\cite{apollov6} runs a red light at an intersection due to tailgating the front car. (Traffic rule violation)}
        \label{fig:case_apollo}
    \end{subfigure}
    \hfill % spacing between the subfigures
    \begin{subfigure}[b]{0.32\linewidth}
        \centering
        \includegraphics[width=\linewidth]{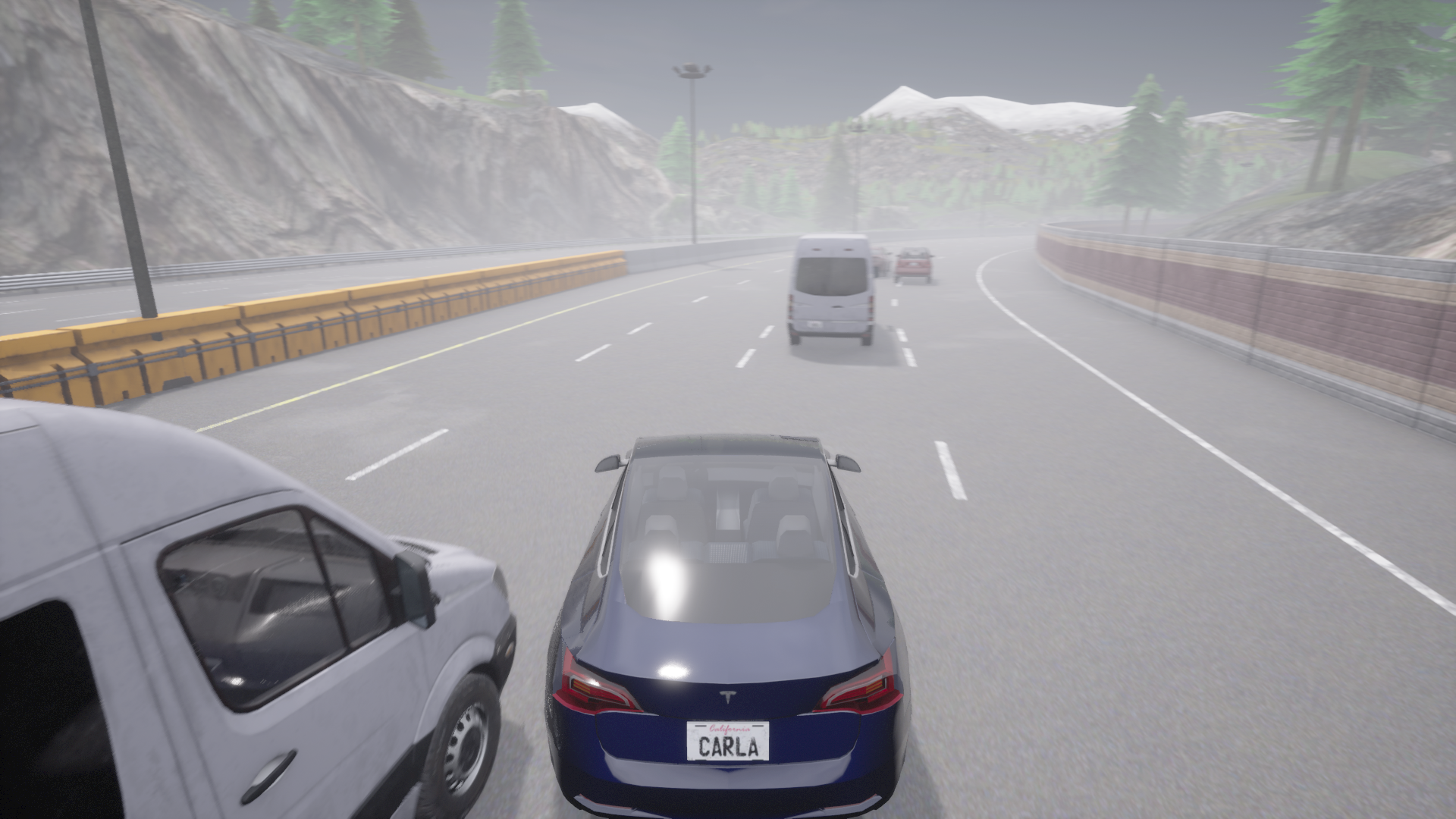}
        \caption{Autoware~\cite{8443742} fails to yield to another vehicle and crashes while merging into the left lane.}
        \label{fig:case_autoware}
    \end{subfigure}
    \hfill % spacing between the subfigures
    \begin{subfigure}[b]{0.32\linewidth}
        \centering
        \includegraphics[width=\linewidth]{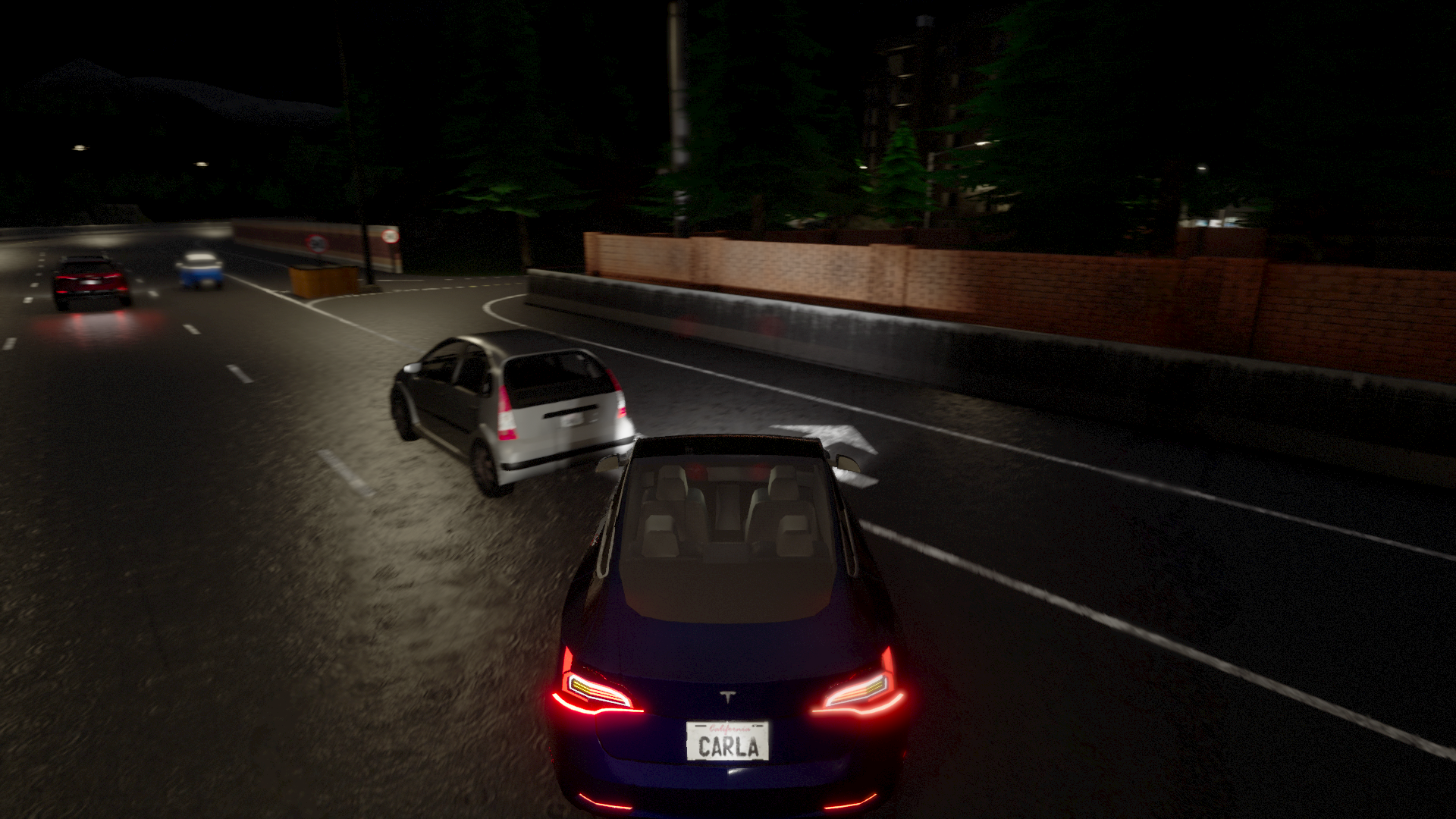}
        \caption{MMFN~\cite{9981775} collides with another vehicle while merging into the right lane to exit the road.}
        \label{fig:case_mmfn}
    \end{subfigure}
    % \hfill

    % Second row with 4 subfigures
    \begin{subfigure}[b]{0.32\linewidth}
        \centering
        \includegraphics[width=\linewidth]{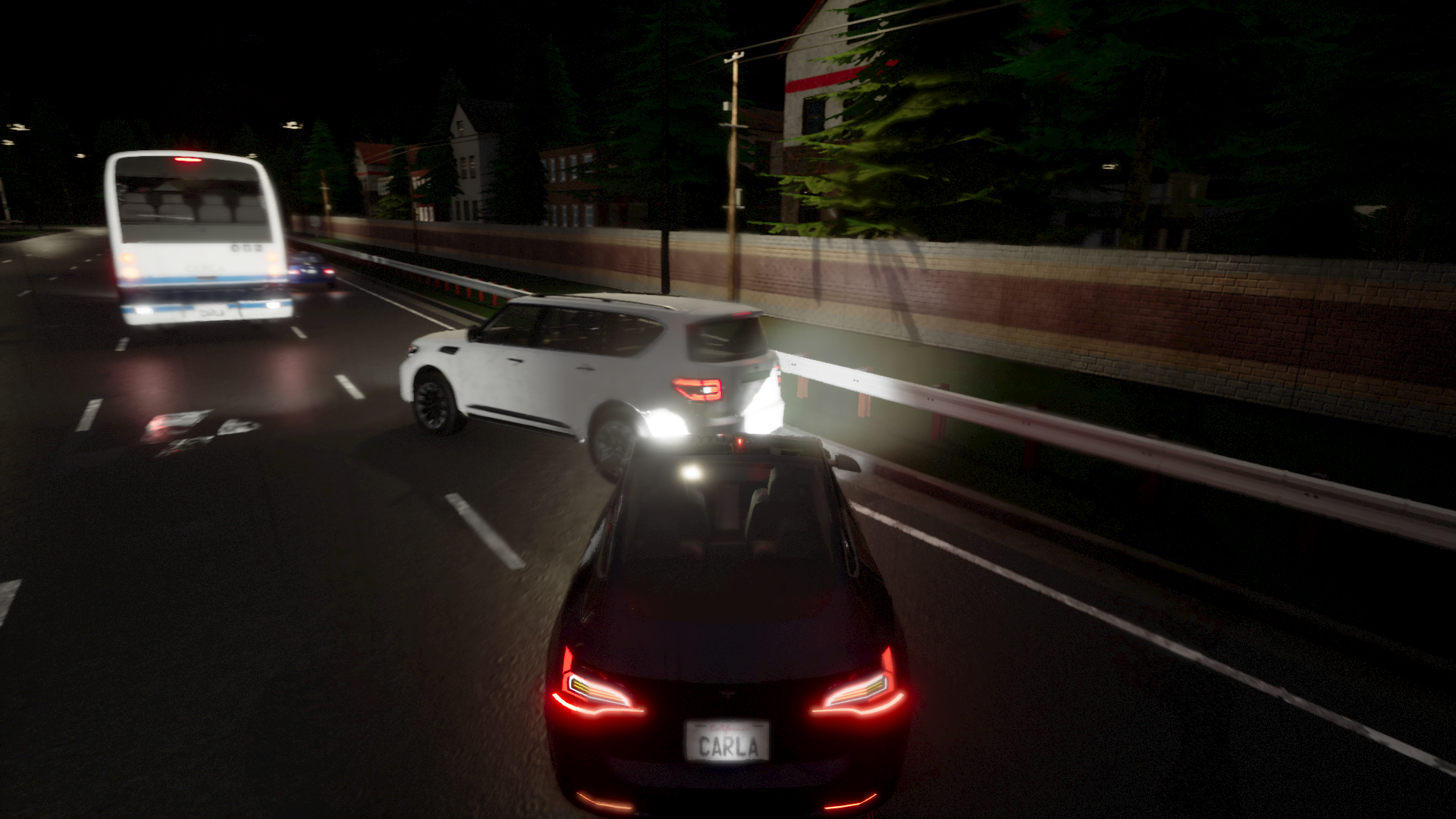}
        \caption{TransFuser~\cite{Chitta2023PAMI} collides with another vehicle while avoiding a stopped car in the rain.}
        \label{fig:case_transfuser}
    \end{subfigure}
    \hfill
    \begin{subfigure}[b]{0.32\linewidth}
        \centering
        \includegraphics[width=\linewidth]{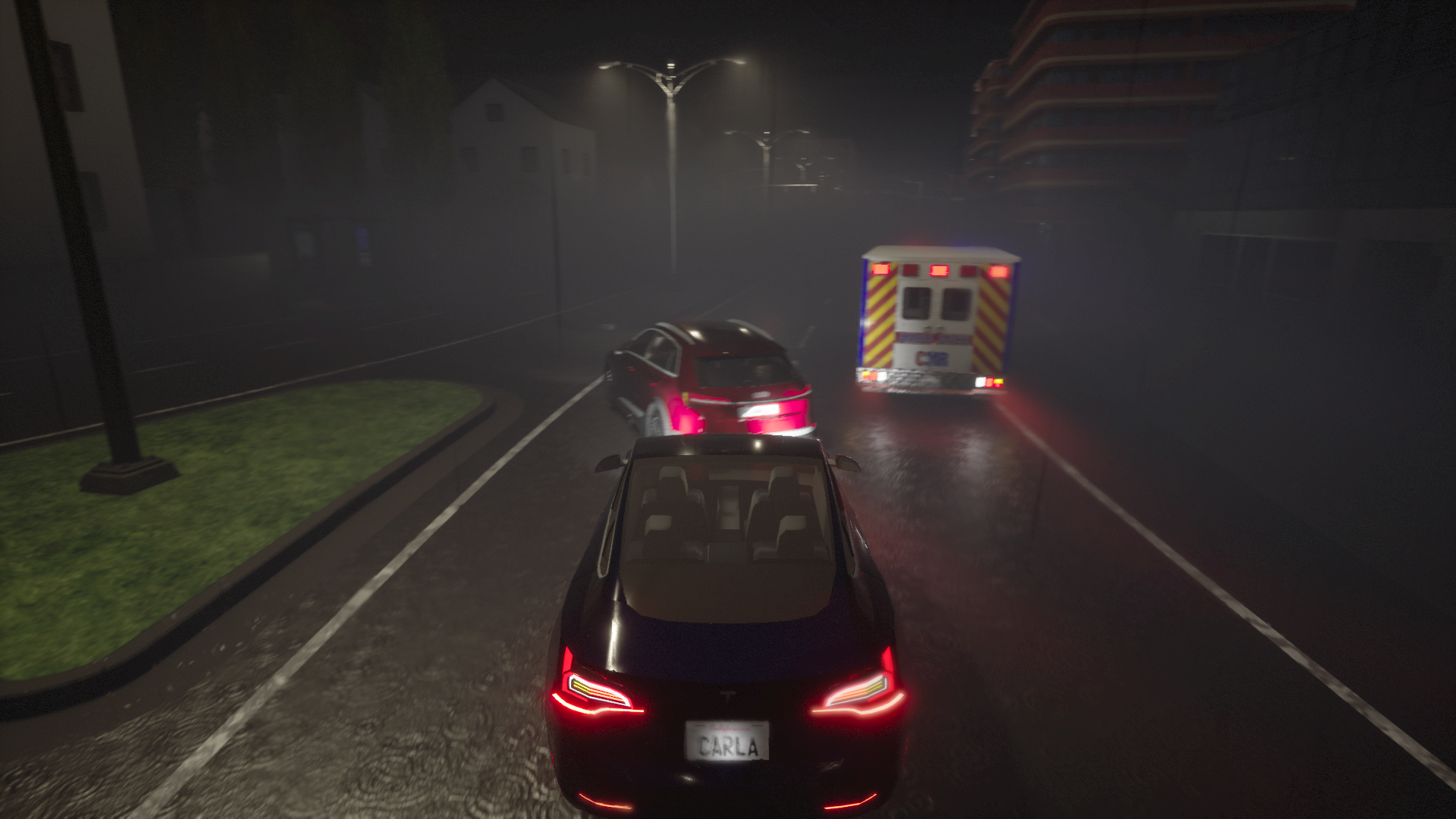}        \caption{Roach~\cite{DBLP:conf/iccv/ZhangLDYG21} collides with the front vehicle while exiting the roundabout due to loss of traction.}
        \label{fig:case_roach}
    \end{subfigure}
    \hfill
    \begin{subfigure}[b]{0.32\linewidth}
        \centering
        \includegraphics[width=\linewidth]{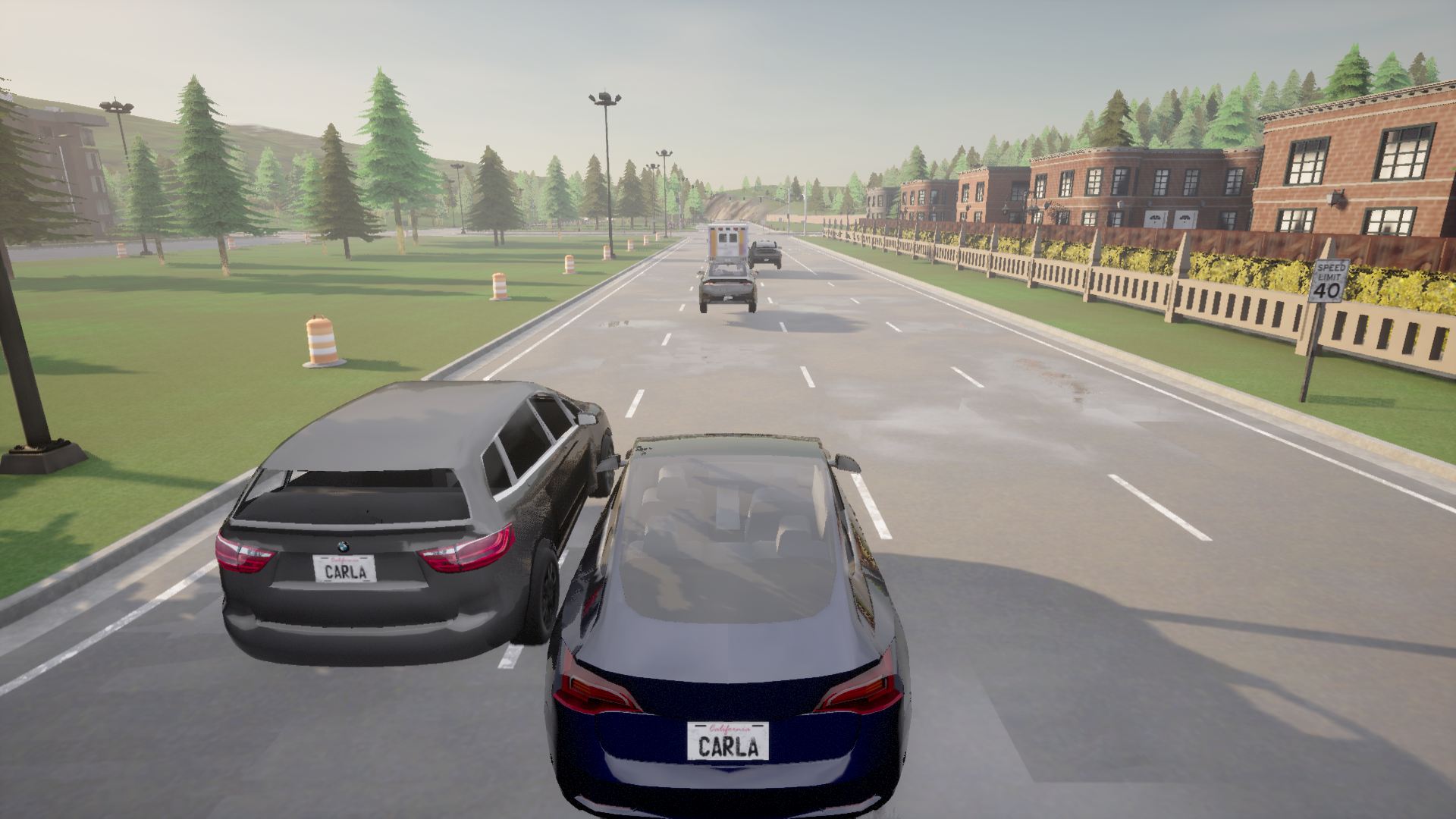}
        \caption{Behavior Agent~\cite{carlaBehaviorAgent} collides with a car merging into the ego vehicle's lane.}
    \end{subfigure}
\end{minipage}
}
\caption{Examples of traffic scenarios that expose collisions or traffic rule violations.}
\label{fig:case}
\end{figure}
% \end{wrapfigure}

%% file: tables/tab_acceff.tex
\begin{table}[hbt]
\centering
\caption{\revisepara{Average number of exposed ADS failures in fuzzing using seed scenarios with different accuracy.}}
\label{tab:acceff}
\resizebox{0.89\columnwidth}{!}{%

\revisepara{
\begin{tabular}{l|c|ccccc}
% \toprule
\specialrule{1pt}{0pt}{0pt}  % Same thickness as \toprule
 & \multicolumn{1}{l|}{LawBreaker} & \multicolumn{5}{c}{DriveFuzz} \\ \cline{2-7} 
Scenario Acc. (Approach) & Apollo & Autoware & MMFN & TransFuser & Roach & Behavior Agent \\ \hline
74.3\% (TARGET) & \result{24.2} & \result{23.6} & \result{21.0} & \result{28.6} & \result{41.4} & \result{41.4} \\
\textbf{97.0\% ({\tool})} & \textbf{\result{39.1}} & \textbf{\result{41.6}} & \textbf{\result{43.8}} & \textbf{\result{45.1}} & \textbf{\result{48.3}} & \textbf{\result{55.2}} \\ 
\specialrule{1pt}{0pt}{0pt}  % Same thickness as \toprule
\end{tabular}%
} % end of revise

}

\end{table}

%% file: tables/tab_info_extract_ablation.tex
\begin{wraptable}{r}{0.3\textwidth}
\centering
\caption{IE accuracy across two modalities in {\tool}.}
\label{tab:ablation}
\resizebox{\linewidth}{!}{%
\begin{tabular}{lc}
\toprule
Modality Used & IE Acc.($\%$) \\ \midrule
Textual Input Only & $79.2$\revisepara{$(\pm 1.1)$} \\
Visual Input Only & $51.8$\revisepara{$(\pm 0.0)$} \\
\textbf{\tool} & \textbf{97.0}\revisepara{$(\pm1.2)$} \\ \bottomrule
\end{tabular}%
}
\end{wraptable}

%% file: tables/tab_info_extract_llm.tex
\begin{wraptable}{r}{0.36\textwidth}
% \begin{wraptable}{r}{0.56\textwidth}
% \begin{table}
\centering
\caption{IE accuracy of {\tool} using different LLMs in the {\textualextractor}.}
\label{tab:llm}
% \resizebox{0.58\linewidth}{!}{%
\resizebox{\linewidth}{!}{%

\begin{tabular}{lc}
\toprule
LLM Used                   & IE Acc.(\%)     \\ 
\midrule
Llama-3.1-8B-Instruct      & $91.1 (\pm1.6)$ \\
Llama-3.1-70B-Instruct     & $92.9 (\pm1.8)$ \\
Llama-3.1-405B-Instruct    & $96.5 (\pm1.5)$ \\
claude-3-5-sonnet-20240620 & $97.0 (\pm2.1)$ \\
mistral-large-2407         & $94.3 (\pm2.0)$ \\
Amazon~Titan~Text~Premier  & $90.7 (\pm1.5)$ \\
gpt-4o-mini-2024-07-18     & $94.0 (\pm1.3)$ \\
gpt-4o-2024-05-13          & $97.0 (\pm1.2)$ \\ 
\bottomrule
\end{tabular}

}
\end{wraptable}
% \end{table}

%% file: tables/tab_info_extract_det+error_det.tex
\begin{table}[htb]
  \centering
  \begin{minipage}{0.22\textwidth}
    \centering

\caption{\revisepara{IE accuracy of {\tool} with errors injected to LLM.}}
\label{tab:error_llm}
\resizebox{0.97\linewidth}{!}{%
\revisepara{
\begin{tabular}{cc}
% \toprule
\specialrule{1pt}{0pt}{0pt}  % Same thickness as \toprule
Error Rate & IE Acc.(\%) \\ 
% \midrule
\hline
1\%        & $95.2 (\pm 1.3)$        \\
2\%        & $95.0 (\pm 1.9)$        \\
3\%        & $93.7 (\pm 2.1)$        \\
4\%        & $93.0 (\pm 2.8)$        \\
5\%        & $92.0 (\pm 2.2)$        \\
6\%        & $89.2 (\pm 2.9)$        \\
7\%        & $88.0 (\pm 1.7)$        \\
8\%        & $87.6 (\pm 1.8)$        \\
9\%        & $85.0 (\pm 3.0)$        \\
10\%       & $84.0 (\pm 2.6)$        \\
% \hline
% Ours      & $97.0(\pm1.2)$        \\ 
% \bottomrule
\specialrule{1pt}{0pt}{0pt}  % Same thickness as \toprule
\end{tabular}%
} % end of revise
} % end of resizebox

  \end{minipage}
  \hfill
  % First minipage for the first table
  \begin{minipage}{0.42\textwidth}
    \centering

\caption{\revisepara{IE accuracy of {\tool} using object detection models in the YOLO family with different model sizes.}}
\label{tab:objdet}
\revisepara{
\resizebox{0.985\linewidth}{!}{%
\begin{tabular}{lccc}
% \toprule
\specialrule{1pt}{0pt}{0pt}  % Same thickness as \toprule
Model     & Release & \# Param.(M)       & IE Acc.(\%) \\ 
% \midrule
\hline
yolov3u   & Mar. 2018    & \phantom{0}N/A  & $94.1(\pm0.8)$   \\
yolov5x6u & June 2020    & 155.4           & $94.2(\pm0.9)$   \\
yolov8x   & Jan. 2024    & \phantom{0}68.2 & $95.0(\pm2.2)$   \\
yolov9e   & Feb. 2024    & \phantom{0}58.1 & $95.4(\pm1.4)$   \\
yolov10n  & May 2024     & \phantom{00}2.3 & $96.1(\pm1.3)$   \\
yolov10s  & May 2024     & \phantom{00}7.2 & $96.2(\pm1.8)$   \\
yolov10m  & May 2024     & \phantom{0}15.4 & $96.9(\pm2.3)$          \\
yolov10b  & May 2024     & \phantom{0}19.1 & $97.0(\pm1.1)$   \\
yolov10l  & May 2024     & \phantom{0}24.4 & $96.8(\pm1.8)$          \\
yolov10x  & May 2024     & \phantom{0}29.5 & $97.0(\pm1.2)$   \\ 
% \bottomrule
\specialrule{1pt}{0pt}{0pt}  % Same thickness as \toprule
\end{tabular}%
} % end of resizebox
} % end of revise

  \end{minipage} 
  % \hspace{0.01\textwidth} % Space between the two tables
  \hfill
  % Second minipage for the second table
  \begin{minipage}{0.31\textwidth}
    \centering

\caption{\revisepara{IE accuracy of {\tool} and object detection precision with error injection.}}
\label{tab:error_det}
\resizebox{\linewidth}{!}{%

\revisepara{
\begin{tabular}{ccc}
% \toprule
\specialrule{1pt}{0pt}{0pt}  % Same thickness as \toprule
Error Rate & Prec.(\%) & IE Acc.(\%) \\ 
% \midrule
\hline
% \specialrule{1pt}{0pt}{0pt}  % Same thickness as \toprule
1\%        & $89.9(\pm 0.4)$       & $92.9 (\pm 1.2)$      \\
2\%        & $88.0(\pm 0.6)$         & $91.7 (\pm 1.4)$     \\
3\%        & $87.8(\pm 0.9)$        & $91.1 (\pm 1.2)$      \\
4\%        & $85.2(\pm 0.7)$         & $89.2 (\pm 1.4)$      \\
5\%        & $83.3(\pm 1.1)$         & $85.9 (\pm 1.6)$      \\
6\%        & $83.0(\pm 0.7)$         & $83.7 (\pm 2.1)$      \\
7\%        & $80.2(\pm 1.0)$         & $81.5 (\pm 1.8)$      \\
8\%        & $77.9(\pm 0.8)$         & $80.0 (\pm 1.5)$      \\
9\%        & $77.6(\pm 0.5)$         & $78.7 (\pm 2.4)$      \\
10\%       & $74.0(\pm 0.7)$         & $76.8 (\pm 1.8)$      \\
% \hline
% Ours      & $98.7(\pm 0.0)$         & $97.0(\pm1.2)$      \\ 
% \bottomrule
\specialrule{1pt}{0pt}{0pt}  % Same thickness as \toprule
\end{tabular}%
} % end of revise
} % end of resizebox

  \end{minipage}
\end{table}

%% file: 4-Discussion.tex
\section{Limitations and Future Work}
\label{sec:discussion}
\textbf{Limitations.}
The visual information extractor in {\tool} leverages two pre-trained Computer Vision models to detect lane divisions and objects such as vehicles, pedestrians, and traffic signals. 
The object detection model, YOLOv10~\cite{wang2024yolov10realtimeendtoendobject}, achieves an average precision of {\accyolo} in our traffic-specific application.
The lane detection model, CLRNet~\cite{DBLP:conf/cvpr/ZhengHLTY0022}, achieves F1 scores of $80.47$ and $96.12$ on two large-scale lane detection benchmarks, CULane~\cite{DBLP:conf/aaai/PanSLWT18} and LLAMAS~\cite{9022318}.
Based on the information extracted by the two models, {\tool} determines \revisecr{the position of each vehicle or pedestrian} using lane occupancy.
This straightforward and transparent algorithm extracts information from the reference image and improves the overall accuracy of the generated traffic scenarios, as shown in Section~\ref{sec:eval_modality}.
However, the effectiveness of {\tool} is inherently tied to the performance of the two computer vision models. 
Despite these models achieving state-of-the-art performance in their respective domains, the accuracy of {\tool}'s visual information extraction may be affected by their detection capabilities.

Moreover, the {\textualextractor} in {\tool} employs an LLM to extract information from textual descriptions, which introduces the potential for hallucination issues inherent in LLMs. 
However, {\tool} mitigates this vulnerability with the {\visualextractor}, which is resistant to such problems.
The {\visualextractor} utilizes a deterministic algorithm to analyze bounding boxes predicted by pre-trained computer vision models, ensuring more reliable and interpretable outputs. 
By aligning \revisecr{the} information from visual and textual input, {\tool} effectively reduces the impact of hallucinations on its overall performance.

\textbf{Future Directions.}
Based on the insights from our evaluation and analysis of the limitations, we propose the following three future directions for traffic scenario generation for ADS testing. 
First, {\tool} currently accepts only reference images as visual input for scenario generation.
Expanding the capabilities of {\visualextractor} to process additional types of visual \revisecr{input}, such as videos or point clouds, could significantly enhance the richness and accuracy of scenario generation.
Second, there is a need for a larger and more diverse set of high-quality traffic scenarios for comprehensive ADS testing.
We plan to extend our benchmark to include more diverse scenarios, which will better serve the needs of the ADS testing community. 
Third, we plan to extend {\tool} to support other DSLs, such as Scenic and OpenSCENARIO, to \revisecr{expand} its potential applications.

\section{Threats to Validity}
\textbf{External validity} 
concerns the generalizability of {\tool}.
One potential threat to validity is the applicability of {\tool} to different ADSs.
While we have made significant efforts to evaluate the effectiveness of {\tool} in ADS testing, it is not feasible to test all available ADSs.
To mitigate this threat, we conduct experiments with a diverse set of {\numads} ADSs, ranging from industrial-level ADSs to simpler ADSs, to assess the bug detection capabilities of {\tool}-generated scenarios.

Another threat to validity is the generalizability of {\tool} to various LLMs.
We mitigate this by evaluating {\tool} with {\numllm} LLMs from {\numllmfamily} model families, including both open-source and closed-source models.
The size of the LLMs ranges from 8B to 405B parameters.
The results demonstrate that {\tool} maintains robust scenario generation capabilities across a wide range of LLMs used in the {\textualextractor}.

\textbf{Internal validity}
concerns the randomness inherent in {\tool} and in the evaluation.
One threat to internal validity is the randomness brought by simulation-based testing.
To mitigate this threat, we repeat all fuzzing experiments ten times to reduce the impact of randomness, \revisecr{and} further conduct statistical analysis to validate the significance of \revisecr{the findings}.
Another potential threat to internal validity is the randomness brought by LLM in {\textualextractor}.
To mitigate this threat, we repeat \revisecr{experiments involving LLMs} {\numrepeatllmexp} times, and report the average results in evaluations.

%% file: 5-Related_Work.tex
\section{Related Work}
\label{sec:related_work}
\subsection{Traffic Scenarios Generation}
The research most related to our proposal is traffic scenario generation approaches~\cite{DBLP:conf/sigsoft/GambiHF19, DBLP:conf/icra/BashettyAF20, fremont2019scenic, zhou2023specification, DBLP:journals/corr/abs-2012-10672, 10234383, 10.1145/3691620.3695037, 10.1145/3597926.3598070, 10.1145/3691620.3695520, tian2024llm}.
%%%% 10234383: from Joshua Garcia
%%%% 10.1145/3691620.3695037: SoVAR
%%%% 10.1145/3597926.3598070: M-CPS Building Critical Testing Scenarios for Autonomous Driving from Real Accidents
%%%% 10.1145/3691620.3695520: LeGEND
%%%% tian2024llm: LEADE
These approaches fall into two categories: (i) reconstructing real-world traffic scenarios from data collected from real-world events,
%% ~\cite{
% DBLP:conf/sigsoft/GambiHF19,  % AC3R 
% 10.1145/3691620.3695037,   % SoVAR
% 10.1145/3691620.3695520,   % LeGEND
% 10.1145/3597926.3598070,  % M-CPS
% DBLP:conf/icra/BashettyAF20}
and (ii) generating traffic scenarios from manually created specifications.

\revisepara{AC3R~\cite{DBLP:conf/sigsoft/GambiHF19} and RMT~\cite{DBLP:journals/corr/abs-2012-10672} use traditional NLP methods
%% the Stanford Core NLP library~\cite{manning-etal-2014-stanford} and a domain-specific ontology to extract information from accident reports.
to extract information from accident reports and traffic rules.
More recently, LLM-based approaches~\cite{deng2023target, 10.1145/3691620.3695520, 10.1145/3691620.3695037} have emerged.
LeGEND~\cite{10.1145/3691620.3695520} introduces a two-stage approach that transforms accident reports into formal logical scenarios using two LLMs
%% and an intermediate representation, thereby 
to generate critical and diverse scenarios. 
SoVAR~\cite{10.1145/3691620.3695037} leverages tailored linguistic patterns in LLM prompts to extract information from accident reports. 
%% It then generates driving trajectories that align with desired road structures by solving a set of driving constraints.
In parallel, video-based methods, DeepCrashTest~\cite{DBLP:conf/icra/BashettyAF20} and M-CPS~\cite{10.1145/3597926.3598070}, reconstruct crash scenarios using vehicle dashboard camera and surveillance camera footage, respectively.}

\revisepara{Compared to existing work, {\tool} introduces a novel multi-modal framework that extracts and aligns complementary information from free-form textual descriptions and visual inputs.
Moreover, while existing methods~\cite{10.1145/3691620.3695520,10.1145/3691620.3695037,deng2023target} typically focus on scenarios with one to two actors in addition to the ego vehicle, {\tool} can generate complex scenarios involving up to \revisecr{$21$} actors.
Additionally, unlike M-CPS~\cite{10.1145/3597926.3598070}, which processes surveillance camera footage using panoptic segmentation, {\tool} leverages images readily available from vehicle dashboard cameras and combines object detection with lane detection to produce a comprehensive description of the traffic scenario.} % end of revise

Both Scenic~\cite{fremont2019scenic} and AVUnit~\cite{zhou2023specification} introduce Domain Specific Languages (DSLs) for specifying traffic scenarios.
However, the complexity of these DSLs presents challenges for ADS developers. Creating large-scale traffic scenarios with them is effortful and time-consuming.
RMT~\cite{DBLP:journals/corr/abs-2012-10672} adopts natural language inputs, offering greater ease of use, but struggles with the inherent ambiguity of natural language that limits detailed scenario description. 
In contrast, {\tool} stands out by leveraging multi-modal inputs for the automated generation of traffic scenarios.
This strategy addresses the challenges posed by natural language ambiguity and DSL complexity and enhances both the precision and efficiency of automated traffic scenario generation.

\subsection{Simulation-based ADS Testing}
In addition to directly generating traffic scenarios, there is also a large body of research focusing on searching for traffic scenarios from initial seeds to test ADSs.
These methods typically start with a set of seed scenarios and adopt various search algorithms to explore the scenario search space.
%%%% DeepRoad - Gordon Fraser
For example, methods like DeepRoad~\cite{DBLP:conf/kbse/ZhangZZ0K18} and DeepTest~\cite{DBLP:conf/icse/TianPJR18} generate mutations based on the original image to test deep neural network-based ADSs.
However, we will not get into details on this topic and focus on simulation-based methods.
%%%% AsFault - Gordon Fraser
AsFault~\cite{DBLP:conf/icse/GambiMF19, DBLP:conf/issta/GambiMF19} searches for complex road network layouts and generates them in BeamNG simulator to test ADS lane-keeping capabilities.
AutoFuzz~\cite{DBLP:journals/tse/ZhongKR23} mutates the starting and destination locations, directions, and velocities of vehicles or pedestrians near the ego vehicle.
Moreover, AV-Fuzzer~\cite{avfuzzer}, EMOOD~\cite{DBLP:conf/kbse/LuoZAJZIW021} and BehAVExplor~\cite{10.1145/3597926.3598072} perform perturbations in driving maneuvers of vehicles or pedestrians around the ego vehicle.
%%%% FITEST - Lionel C. Briand
FITEST~\cite{DBLP:conf/kbse/AbdessalemPNBS18}, LawBreaker~\cite{DBLP:conf/kbse/0008P00Y22}, and DriveFuzz~\cite{DBLP:conf/ccs/KimLRJ0K22} further expand the scope by mutating more factors of traffic scenarios, such as weather and traffic signals. 
% metamorphic testing~\cite{DBLP:conf/icse/TianPJR18, DBLP:conf/kbse/ZhangZZ0K18}, 
% adversarial testing~\cite{DBLP:conf/icse/ZhouLKGZ0Z020}, ontology-based testing~\cite{8539174, LI2020106200, 8728928}, etc.
% Please refer to Lou et al.'s recent survey for more details~\cite{10.1145/3540250.3549111}
ConfVE~\cite{10.1145/3660792} focuses on searching the ADS configuration space to detect unique system failures.
\revisepara{LEADE~\cite{tian2024llm} proposes to model evolutionary search as a question-answering task and to use LLMs to search for diverse scenarios.}
Our proposed method is complementary to these methods since the traffic scenarios generated by our method can serve as high-quality initial seeds, as demonstrated in our comprehensive experiments.

%% file: 6-Conclusion.tex
\section{Conclusion}
This paper presents {\tool}, a novel multi-modal traffic scenario generator specifically designed for ADS testing.
The primary innovation of {\tool} lies in its unique integration of textual and visual inputs from ADS developers, leveraging the clarity and abstraction of textual descriptions and the detailed specificity of visual inputs.
Evaluated on a benchmark of {\numscenario} traffic scenarios, {\tool} achieves \result{$97.0\%$} accuracy, outperforming the best-performing baseline by \result{$7.3\%$}. 
Testing ADSs directly using these scenarios successfully identifies {\numdirectbug} bugs.
As initial seeds in fuzzing, these scenarios significantly improve bug detection effectiveness by {\ratiobug} and time efficiency by {\ratiotime}.
An ablation study demonstrates the contribution of each input modality, while further experiments confirm the generalizability of {\tool} across different large language models \revisecr{and computer vision models}.
% The broader impact of our research is evident in the versatility and applicability of {\tool} across various contexts. 
Its ability to seamlessly adapt to different traffic scenarios, ADS configurations, and fuzzing algorithms highlights its potential as a transformative framework in the field of ADS testing.

\section{Data Availability}
\revisecr{The source code and data are publicly available for replication and reproduction~\cite{my_artifacts}.}